\documentclass[a4paper,fleqn]{cas-dc}

\usepackage[numbers]{natbib}

\usepackage{booktabs}
\usepackage[english]{babel} 
\usepackage{url}
\usepackage{pifont}
\usepackage{float}
\usepackage{paralist}
\usepackage{enumitem}
\usepackage{multirow}
\usepackage{menukeys}
\usepackage{romannum}
\usepackage{comment}
\usepackage{multicol}
\usepackage{rotating}
\usepackage{microtype}
\usepackage{mdframed}
\usepackage{csquotes}
\usepackage{bm}
\usepackage[listings,skins,breakable]{tcolorbox}
\usepackage{environ}
\usepackage{pdflscape}

\DeclareUrlCommand\doi{\def\UrlLeft##1\UrlRight{DOI\nobreakspace\href{http://dx.doi.org/##1}{##1}}\urlstyle{rm}} 

\newcommand*\circled[1]{\tikz[baseline=(char.base)]{
            \node[shape=circle,fill,inner sep=0.5pt] (char) {\textcolor{white}{#1}};}}
            
\renewcommand\vec[1]{\overrightarrow{#1}}

\newcommand\cev[1]{\overleftarrow{#1}}

\NewEnviron{myequation}{%
\begin{equation}
\scalebox{0.9}{$\BODY$}
\end{equation}
}

\begin{document}
\let\WriteBookmarks\relax
\def\floatpagepagefraction{1}
\def\textpagefraction{.001}

\shorttitle{Automatic Creation of Acceptance Tests by Extracting Conditionals from Requirements}

\shortauthors{Fischbach et~al.}

\title [mode = title]{Automatic Creation of Acceptance Tests by Extracting Conditionals from Requirements: NLP Approach and Case Study}                   


%
\author[1,2]{Jannik Fischbach}[orcid=0000-0002-4361-6118]

\cormark[1]
\cortext[cor1]{Corresponding author}
\ead{jannik.fischbach@netlight.com}
\credit{Conceptualization, Methodology, Software, Validation, Investigation, Data Curation, Writing - Original Draft, Writing - Review \& Editing, Visualization, Supervision}

\address[1]{Netlight Consulting GmbH, Sternstraße 5, Munich, 80538, Germany}

\address[2]{fortiss GmbH, Guerickestraße 25, Munich, 80805, Germany} 

\author[3]{Julian Frattini}[orcid=0000-0003-3995-6125]
\ead{julian.frattini@bth.se}
\credit{Methodology, Software, Validation, Investigation, Data Curation, Writing - Review \& Editing, Visualization}

\address[3]{Blekinge Institute of Technology, Valhallavägen 1, 371 41, Karlskrona, Sweden}

\author[4]{Andreas Vogelsang}[orcid=0000-0003-1041-0815]
\ead{vogelsang@cs.uni-koeln.de}
\credit{Conceptualization, Supervision, Project administration}

\address[4]{University of Cologne, Albertus-Magnus-Platz, 50923, Cologne, Germany}
    
\author[2,3]{Daniel Mendez}[orcid=0000-0003-0619-6027]
\ead{daniel.mendez@bth.se}
\credit{Methodology, Writing - Review \& Editing, Supervision}

\author[3]{Michael Unterkalmsteiner}[orcid=0000-0003-4118-0952]
\ead{michael.unterkalmsteiner@bth.se}
\credit{Methodology, Writing - Review \& Editing, Supervision}

\author[6]{Andreas Wehrle}
\ead{andreas.wehrle@allianz.de}
\credit{Data curation, Software}

\author[1]{Pablo Restrepo Henao}
\ead{pablo.restrepo@netlight.com}
\credit{Software}

\author[5]{Parisa Yousefi}
\ead{parisa.yousefi@ericsson.com}
\credit{Resources, Investigation}

\author[5]{Tedi Juricic}
\ead{tedi.juricic@ericsson.com}
\credit{Resources, Investigation}

\address[5]{Ericsson, Ölandsgatan 1, 371 33, Karlskrona, Sweden}
    
\author[6]{Jeannette Radduenz}
\ead{jeannette.radduenz@allianz.de}
\credit{Resources, Investigation}

\address[6]{Allianz Deutschland AG, Dieselstr. 6, 85774, Unterföhring, Germany}
    
\author[7]{Carsten Wiecher}
\ead{c.wiecher@kostal.com}
\credit{Resources, Investigation}

\address[7]{Leopold Kostal GmbH \& Co. KG, An der Bellmerei 10, Lüdenscheid, 58513, Germany}    
    
\begin{abstract}
Acceptance testing is crucial to determine whether a system fulfills end-user requirements. However, the creation of acceptance tests is a laborious task entailing two major challenges: (1) practitioners need to determine the right set of test cases that fully covers a requirement, and (2) they need to create test cases manually due to insufficient tool support. Existing approaches for automatically deriving test cases require semi-formal or even formal notations of requirements, though unrestricted natural language is prevalent in practice. In this paper, we present our tool-supported approach \textsf{CiRA (\textbf{C}onditionals \textbf{i}n \textbf{R}equirements \textbf{A}rtifacts)} capable of creating the minimal set of required test cases from conditional statements in informal requirements. We demonstrate the feasibility of \textsf{CiRA} in a case study with three industry partners. In our study, out of 578 manually created test cases, 71.8~\% can be generated automatically. Additionally, \textsf{CiRA} discovered 80 relevant test cases that were missed in manual test case design. \textsf{CiRA} is publicly available at \url{www.cira.bth.se/demo/}.
\end{abstract}




\maketitle

\section{Introduction}
\label{sec:intro}

Acceptance tests are used to verify the conformity between end-user requirements and actual system behavior~\cite{ISONorm}. Each acceptance test contains a finite set of test cases that specify certain test inputs and expected results. Test case design is a very laborious activity that easily accounts for 40~-~70~\% of the total effort in the testing process~\cite{beller15}. This stems from the following challenges. 

\paragraph{Challenge 1} Determining the right set of test cases that fully covers a requirement is a difficult task, especially for complex requirements. A requirement is considered fully covered if a set of associated test cases assures the behavior implied by that requirement~\cite{whalen06}. In a previous study~\cite{fischbachESEM}, we found that acceptance tests are often not systematically created, resulting in incomplete or excessive test cases. In the case of missing test cases, system defects are not (or only partially) detected. In contrast, excessive test cases lead to unnecessary testing efforts and increased test maintenance costs. Consequently, practitioners need to strike a balance between full test coverage and number of required test cases.

\paragraph{Challenge 2} Creating acceptance tests is a predominantly manual task due to insufficient tool support~\cite{GAROUSI}. Most of the existing approaches allow the derivation of test cases from semi-formal requirements~\cite{wangbriand,Carvalho14,barros11} (e.g., expressed in controlled natural language) or formal requirements~\cite{shaoying,enase14} (e.g., expressed in linear temporal logic), but are not suitable to process informal requirements. However, studies~\cite{Kassab14} have shown that requirements are usually expressed in unrestricted natural language (NL). Some approaches~\cite{fischbachICST,Verma,santiagoSpace} address this research gap and focus on deriving test cases from informal requirements. Nevertheless, they show poor performance when evaluated on unseen real-world data. Specifically, they are not robust against grammatical errors and fail to process words that are not yet part of their training vocabulary~\cite{fischbachAIRE21}.

\paragraph{Research Goal} We aim to develop a tool-supported approach to derive the minimal set of required test cases automatically from NL requirements by applying \textit{Natural Language Processing}~(NLP).

\paragraph{Principal Idea} Functional requirements often describe system behavior by relating events to each other, e.g., \enquote{If the system detects an error ($e_1$), an error message shall be shown ($e_2$)}. Previous studies~\cite{fischbachREFSQ,fischbachICST} show that such conditional statements are prevalent in both traditional and agile requirements such as acceptance criteria. In this paper, we focus on conditionals in NL requirements and utilize their embedded logical knowledge for the automatic derivation of test cases. We answer three research questions (RQ):

\begin{itemize}[]
  \item \textbf{RQ 1}: How to extract conditionals from NL requirements and use their implied relationships for automatic test case derivation?
  \item \textbf{RQ 2}: Can our automated approach create the same test cases as the manual approach?
  \item \textbf{RQ 3}: What are the reasons for deviating test cases?
\end{itemize}

The answers to RQ 1 shall inform the implementation of a new tool-supported approach for automatic test case derivation. RQ 2 and RQ 3 study the impact of the new approach: does it achieve the status quo or even lead to an improvement of the manual test case derivation? To this end, we conduct a case study with three industry partners and compare automatically created test cases with existing, manually created test cases. In summary, this paper makes the following contributions (C):

\begin{itemize}[]
  \item \textbf{C 1}: To answer RQ 1, we present our tool-supported approach \textsf{CiRA} (\textbf{C}onditionals \textbf{i}n \textbf{R}equirements \textbf{A}rtifacts) capable of (1) detecting conditional statements in NL requirements, (2) extracting their implied relationships in fine-grained form and (3) mapping these relationships to a \textit{Cause-Effect-Graph} from which the minimal set of required test cases can be derived automatically. The output of \textsf{CiRA} are manual test cases, which - if required - can be converted into automatic test cases using third-party tools such as Selenium, Robot Framework, or Tricentis.
  \item \textbf{C 2}: To answer RQ 2 and RQ 3, we conduct a case study with three companies and compare \textsf{CiRA} to the manual test case design. We show that \textsf{CiRA} is able to automatically generate 71.8~\% of the 578 manually created test cases. In addition, \textsf{CiRA} identifies 80 relevant test cases that were missed in the manual test case design.
  \item \textbf{C 3}: To strengthen transparency and facilitate replication, we make our tool, code, annotated data set, and all trained models publicly available.\footnote{A demo of \textsf{CiRA} can be accessed at \url{http://www.cira.bth.se/demo/}. Our code, annotated data sets, and all trained models can be found at \doi{10.5281/zenodo.5550387}.}
\end{itemize}

\paragraph{Running Example}
In the course of the paper, we demonstrate the functionality of \textsf{CiRA} by means of a running example. Specifically, we explain how \textsf{CiRA} automatically derives the minimum number of required test cases for the requirements specification shown in \autoref{fig:runningExample}. The specification contains an excerpt of requirements that describe the functionality of \enquote{The Energy Management System} (THEMAS). THEMAS is intended to be used by people that maintain the heating and cooling systems in a building. We retrieved the requirements from the PURE (PUblic REquirements) data set~\cite{Ferrari17} that contains 79 publicly available natural language requirements documents collected from the Web. We encourage the readers of this paper to use our online demo to process the running example on their own, allowing them to follow each individual step of \textsf{CiRA}.

\begin{figure}
\small
\begin{mdframed}
    \begin{itemize}[leftmargin=1em,labelwidth=*,align=left]
    
\item \textbf{REQ A:} If the temperature change is requested, then the determine heating/cooling mode process is activated and makes a heating/cooling request.

\item \textbf{REQ B:} If the current temperature value is strictly less than the lower value of the valid temperature range or if the received temperature value is strictly greater than the upper value of the valid temperature range, then the THEMAS system shall identify the current temperature value as an invalid temperature and shall output an invalid temperature status.

\item \textbf{REQ C:} The THEMAS system shall maintain the ON/OFF status of each heating and cooling unit.

\item \textbf{REQ D:} Temperatures that do not exceed these limits shall be output for subsequent processing.

\item \textbf{REQ E:}
If this condition is true, then this module shall output a request to turn on the heating unit in case LO = T LT.

\item \textbf{REQ F:}
The heating/cooling unit shall have no real-time delay when these statuses are sent to the THEMAS system.

\item \textbf{REQ G:}
Each thermostat shall have a unique identifier by which that thermostat is identified in the THEMAS system.

\item \textbf{REQ H:}
When an event occurs, the THEMAS system shall identify the event type and format an appropriate event message.
\end{itemize}
\end{mdframed}
\caption{Requirements specification of THEMAS~\cite{Ferrari17}}
\label{fig:runningExample}
\end{figure}

\paragraph{Outline} The remainder of this paper is organized as follows: \autoref{sec:fundamentals} provides the theoretical background. \autoref{sec:approach} answers RQ~1 and introduces \textsf{CiRA} in detail. \autoref{sec:evaluation} presents the results of our case study and answers RQ~2 and RQ~3. \autoref{sec:discussion} discusses our results and indicates directions for both research and practice. \autoref{sec:related} briefly surveys related work. Finally, \autoref{sec:conclusion} presents our conclusions.

\begin{figure*}
    \centering
    \includegraphics[trim={0.5cm 1cm 0.5cm 0}, width=\textwidth]{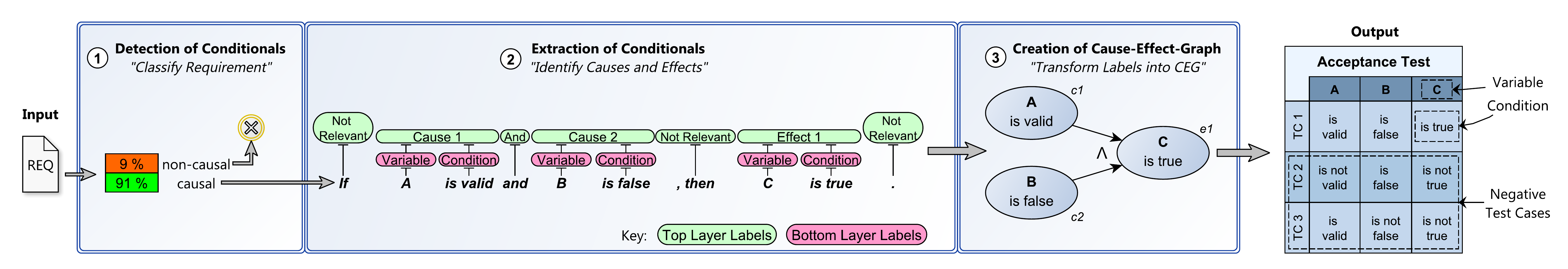}
    \caption{Overview of the \textsf{CiRA} pipeline consisting of three steps: (1) detection of conditionals, (2) fine-grained extraction of conditionals, and (3) CEG creation. Processed REQ: \textit{If A is valid and B is false, then C is true.}}
    \label{approach_overview}
\end{figure*}

\section{Fundamentals}
\label{sec:fundamentals}

\paragraph{Test Case} A test case is a set of certain test inputs (input parameters) and expected results (output parameters) used to verify compliance with a specific requirement~\cite{ISONorm}. Each input and output parameter is defined by a variable and a condition that the parameter can take~\cite{Sneed07}. For example, the parameter \enquote{the system detects an error} can be decomposed into \keys{\textbf{Variable}: the system} and \keys{\textbf{Condition}: detects an error}. All test cases that constitute a single acceptance test are summarized in a test case specification (see \autoref{approach_overview}, right). Each row represents a test case. The variables of the input and output parameters are listed in the columns. The conditions of the parameters that shall be inspected as part of a certain test case are contained in the respective cells. 

\paragraph{Conditional Statements} A conditional statement (short: conditional) is a grammatical structure consisting of two parts: an adverbial clause, often referred to as the \textit{antecedent}, and a main clause, also known as the \textit{consequent}. This can be illustrated by the following REQ: 
\begin{equation*}\label{eq:exampleConditional}
\small
\text{If } \underbrace{\text{the system detects an error}}_\text{antecedent} \text{ , } \underbrace{\text{an error message shall be shown}}_\text{consequent} \text{ . }
\end{equation*}

The relationship between an antecedent and consequent can be interpreted logically in two different ways~\cite{fischbachRENEXT}. First, by means of an implication as $e_1 \Rightarrow e_2$, in which the antecedent is a \textit{sufficient} condition for the consequent. Interpreting REQ as an implication requires the system to display an error message if the antecedent is true. However, it is not specified what the system should do if the antecedent is false. The implication allows both the occurrence of the consequent and its absence if the antecedent is false. The case of $e_1 = $ false is \emph{underspecified}. From a testing point of view, underspecified requirements can be problematic because the negative scenario is not defined. In fact, when reading REQ it may as well be reasonable to assume that the error message shall not be shown if the error has not been detected. This interpretation corresponds to a logical equivalence ($e_1 \Leftrightarrow e_2$), where the antecedent is both a \textit{sufficient} and \textit{necessary} condition for the consequent. Interpreting REQ as an equivalence requires the system to display an error message if and only if it detects an error. 

In a previous study~\cite{fischbach2021profes}, we found that conditionals in requirements are interpreted ambiguously and that practitioners disagree whether antecedents are only \textit{sufficient} or also \textit{necessary} for the consequent. To ensure that the automatically derived test cases correspond to the different logical interpretations, we require two variants of test case generation: The first variant interprets conditionals as implications and generates only the positive test cases. The second variant interprets the conditionals as equivalences and generates both the positive and negative test cases. \textsf{CiRA} supports both variants of test case generation. The user can choose whether s/he perceives antecedents to be both \textit{sufficient} and \textit{necessary} conditions for consequents or not. Depending on the selection, we filter the derived test cases and display the acceptance test that corresponds to the user's interpretation. In the given example (see \autoref{approach_overview}), we perceive the antecedents as \textit{necessary} conditions for the consequent. Accordingly, our approach created one positive test case (see TC 1) and two negative test cases (see TC 2 and TC 3). 

For the sake of readability, in the remainder of the paper, we denote antecedents as causes and consequents as effects. Additionally, we term sentences that contain conditionals as causal sentences.

\paragraph{Cause Effect Graph} A \textit{Cause-Effect-Graph}~(CEG) can be interpreted as a combinatorial logic network, which describes the interaction of causes and effects by Boolean logic~\cite{Nursimulu95}. It consists of nodes for each cause and effect and uses arcs with Boolean operators (conjunction $\wedge$, disjunction $\vee$, negation $\neg$) to illustrate the relationship between the nodes. Let $G$ be the CEG shown in \autoref{approach_overview} with effect set $E$ and cause set $C$. In the example, $|C| = 2$ including $c1$, and $c2$ while $|E| = 1$ with $e1$. To derive test cases from $G$, the \textit{Basic Path Sensitization Technique}~(BPST) is applied. The graph is traversed back from the effects to the causes and test cases are created according to specific decision rules (cf.~\cite{Myers} and~\cite{Nursimulu95}). These rules achieve the maximum probability of finding failures while avoiding the complexity of generating 2\textsuperscript{n} test cases, where $n$ is the number of causes. Hence, \textit{Cause-Effect-Graphing} allows us to support practitioners in balancing between sufficient test coverage and the lowest possible number of test cases. This is also indicated by the acceptance test in \autoref{approach_overview}, which contains the test cases generated by BPST for $G$. To check the functionality presented in the CEG comprehensively, only 3 test cases are needed instead of the maximum number of 2\textsuperscript{3} test cases.

\section{\textsf{CiRA} Pipeline}
\label{sec:approach}
As shown in \autoref{approach_overview}, \textsf{CiRA} consists of three steps: We first detect whether an NL requirement contains a conditional (see \autoref{sec:detection}). Second, we extract the conditional in fine-grained form (see \autoref{sec:extraction}). Specifically, we consider the combinatorics between causes and effects and split them into more granular text fragments (e.g., variable and condition), making the extracted conditionals suitable for automatic test case derivation. Third, we map the extracted causes and effects into a CEG to derive the minimum number of required test cases (see \autoref{sec:creation}).

\paragraph{Input Representation} \textsf{CiRA} is not dependent on any specific format of requirements. Rather, it is able to process any kind of NL representation (i.e., \textsf{CiRA} does not dictate the semantics and syntax that a requirements author must follow). Hence, it understands unstructured representations (e.g., If event A occurs, then event B evaluates to true) as well as semi-structured formulations (e.g., IF A is true AND B is false THEN the system shall shut down). However, \textsf{CiRA} is not trained to process requirements that span multiple sentences.

\subsection{Detection of Conditionals}\label{sec:detection}

\paragraph{Problem} We define the detection of conditionals in NL sentences as a binary classification problem, in which we are given a certain NL sentence $\mathcal{X}$ and we are required to produce a nominal label $y \in \mathcal{Y} = \{\text{causal}, \text{non-causal}\}$.

\paragraph{Novelty} 
In previous work~\cite{fischbachREFSQ}, we compared the performance of different approaches for this task: baseline systems that search for cue phrases (e.g., \textit{if}, \textit{when}) that usually indicate causes and effects, Machine Learning (ML) approaches (e.g., Random Forest) and Deep Learning (DL) approaches (e.g., BERT+Softmax). In this paper, we present only the best-performing method that combines syntactically enriched BERT embeddings with a softmax classifier. For a detailed comparison of all investigated approaches, please refer to our manuscript~\cite{fischbachREFSQ}. Our detection approach is trained on 8,430 NL requirements of which 4,215 sentences are causal. It achieves a \textit{macro}-$F_{1}$ score of 82~\% and outperforms related approaches with an average gain of 11.06~\% in \textit{macro}-Recall and 11.43~\% in \textit{macro}-Precision.

\begin{figure}
  \centering
  \includegraphics[trim={0 1cm 0 0},width=\linewidth]{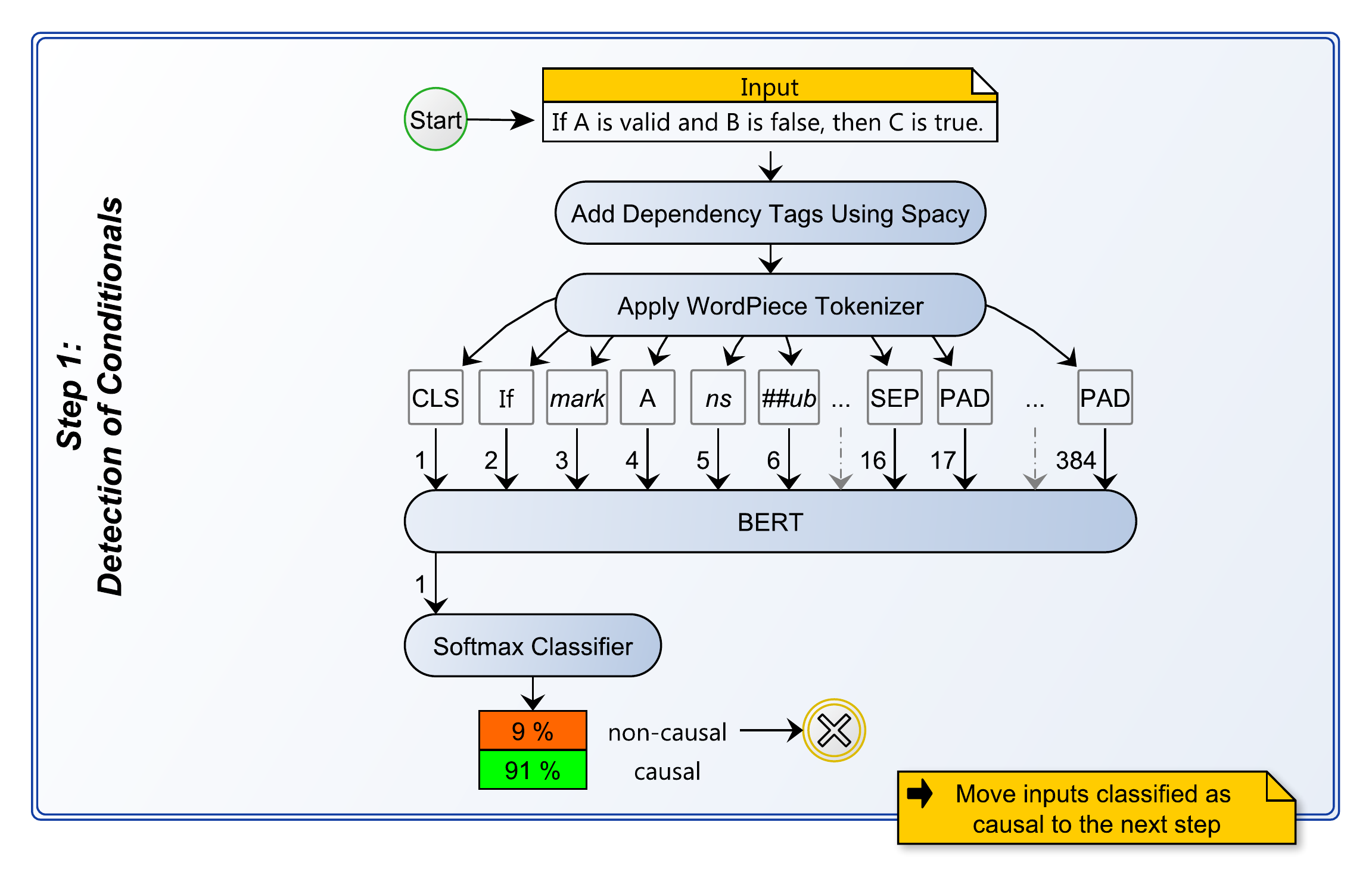}
  \caption{In-depth visualization of the first step in the \textsf{CiRA} pipeline: the detection of conditionals in NL requirements. Processed REQ: \textit{If A is valid and B is false, then C is true.}}
  \label{nurdetection}
\end{figure}

\paragraph{Solution} Our detection algorithm consists of two layers: 1) embedding layer and 2) inference layer. For a detailed visualization of the internal behavior of the individual layers, please refer to \autoref{nurdetection}.

\textbf{1) Embedding Layer} We represent each sentence as a sequence of word embeddings. Let us denote $s$ as a sentence with $n$ tokens: $s = \{v_1,v_2,\dots,v_n\}$, where vector $v_i$ represents the vector of $i$-th token with a dimension of $d$. In recent years, several methods have been developed to implement word embeddings. Traditional methods like word2vec~\cite{mikolov13} are capable of transforming a word into a single vector representation. However, they do not consider the context of the word. Hence, words are always represented as the same vector, although they can have different meanings depending on their context. To address this problem, contextual word embeddings like Bidirectional Encoder Representations from Transformers (BERT)~\cite{bert} were developed. Since BERT outperforms its predecessors, ELMo and GPT-2, in a number of NLP tasks, we use BERT embeddings in our proposed architecture. BERT requires input sequences with a fixed length. Therefore, sentences that are shorter are adjusted to this fixed length by appending padding tokens (PAD). Other tokens such as the separator (SEP) and classification (CLS) token, are also inserted to provide further information about the sentence to the model, where SEP marks the end of a sentence. CLS is the first token in the sequence and represents the whole sentence (i.e., it is the pooled output of all tokens of a sentence). Studies~\cite{sundararaman2019} have shown that the performance of NLP models can be improved by providing explicit prior knowledge of syntactic information to the model. Hence, we enrich the input sequence by adding the corresponding dependency (DEP) tag to each token and feed it into BERT. To choose a suitable fixed length for our input sequences, we analyzed the lengths of the sentences in our data set. A length of 384 tokens showed to be reasonable and allowed us to keep BERT’s computational requirements to a minimum.

\textbf{2) Inference Layer} For our classification task, we only use the CLS token because it stores the information of the whole sentence. We feed the pooled information into a single-layer feedforward neural network that uses a softmax layer, which calculates the probability that a sentence contains a conditional or not: $\hat y = \text{softmax}(Wv_0 + b)$, where $\hat y$ is the predicted label for the sentence, $W$ is the weighted matrix, $v_0$ is the first token in the sentence (i.e., the CLS token), and $b$ is the bias. We select the class with the highest probability as the final classification result.

\paragraph{Running Example}
When applying \textsf{CiRA} to the requirements specification shown in \autoref{fig:runningExample}, \textsf{CiRA} performs the first step of its pipeline: it tries to identify which requirements in the specification contain a conditional. For this purpose, REQ A - H are tokenized in the embedding layer and enriched with dependency tags as described in the previous section. Finally, the CLS token of each REQ is passed to the inference layer, where a softmax layer computes the probability of the REQ being causal or not. In the present example, \textsf{CiRA} classifies REQ A, REQ B, REQ D, REQ E, REQ F, and REQ H as causal and correctly discovers that REQ C and REQ G do not contain a conditional statement. Hence, REQ C and REQ G are excluded from the further test generation process. The remaining requirements are forwarded to the next step, namely the extraction of conditionals.

\begin{table*}
\footnotesize
\caption{Overview of the class distribution (sentence and token level) in our training, validation and testing data sets. Annotation validity per class is reported as pair-wise averaged $F_{1}$ score.}\label{tab:dataanalysis}
\resizebox{\textwidth}{!}{\begin{tabular}{@{}clllllllllllllc@{}}
\toprule
\multicolumn{1}{l}{}                                                             &                                                & \multicolumn{3}{c}{\textbf{Complete Dataset}}                                                                                                                                            & \multicolumn{3}{c}{\textbf{Training Set}}                                                                                                                                                 & \multicolumn{3}{c}{\textbf{Validation Set}}                                                                                                                                             & \multicolumn{3}{c}{\textbf{Testing Set}}                                                                                                                                                &                                                                                   \\ \cmidrule(lr){3-14}
\multicolumn{1}{l}{}                                                             & \textbf{Label Type}                            & Sentences                    & \multicolumn{1}{c}{\begin{tabular}[c]{@{}c@{}}WordPiece \\ Tokens\end{tabular}} & \multicolumn{1}{c}{\begin{tabular}[c]{@{}c@{}}BPE\\ Tokens\end{tabular}} & Sentences                    & \multicolumn{1}{c}{\begin{tabular}[c]{@{}c@{}}WordPiece \\ Tokens\end{tabular}} & \multicolumn{1}{c}{\begin{tabular}[c]{@{}c@{}}BPE \\ Tokens\end{tabular}} & Sentences                   & \multicolumn{1}{c}{\begin{tabular}[c]{@{}c@{}}WordPiece \\ Tokens\end{tabular}} & \multicolumn{1}{c}{\begin{tabular}[c]{@{}c@{}}BPE\\ Tokens\end{tabular}} & Sentences                   & \multicolumn{1}{c}{\begin{tabular}[c]{@{}c@{}}WordPiece\\ Tokens\end{tabular}} & \multicolumn{1}{c}{\begin{tabular}[c]{@{}c@{}}BPE \\ Tokens\end{tabular}} & \multicolumn{1}{c}{\begin{tabular}[c]{@{}c@{}}Annotation\\ Validity\end{tabular}} \\ \midrule
                                                                                 & \cellcolor[HTML]{EFEFEF}\textbf{Cause 1}      & \cellcolor[HTML]{EFEFEF}1946 & \cellcolor[HTML]{EFEFEF}18743                                              & \cellcolor[HTML]{EFEFEF}18317                                                & \cellcolor[HTML]{EFEFEF}1556 & \cellcolor[HTML]{EFEFEF}14862                                              & \cellcolor[HTML]{EFEFEF}14499                                                 & \cellcolor[HTML]{EFEFEF}194 & \cellcolor[HTML]{EFEFEF}1878                                               & \cellcolor[HTML]{EFEFEF}1848                                                 & \cellcolor[HTML]{EFEFEF}196 & \cellcolor[HTML]{EFEFEF}2003                                              & \cellcolor[HTML]{EFEFEF}1970                                                  & \cellcolor[HTML]{EFEFEF}87~\%                                                      \\
                                                                                 & \textbf{Cause 2}                              & 661                          & 5158                                                                       & 5190                                                                         & 523                          & 4036                                                                       & 4072                                                                          & 68                          & 540                                                                        & 534                                                                          & 70                          & 582                                                                       & 584                                                                           & 71~\%                                                                              \\
                                                                                 & \cellcolor[HTML]{EFEFEF}\textbf{Cause 3}      & \cellcolor[HTML]{EFEFEF}137  & \cellcolor[HTML]{EFEFEF}1109                                               & \cellcolor[HTML]{EFEFEF}1102                                                 & \cellcolor[HTML]{EFEFEF}105  & \cellcolor[HTML]{EFEFEF}856                                                & \cellcolor[HTML]{EFEFEF}853                                                   & \cellcolor[HTML]{EFEFEF}18  & \cellcolor[HTML]{EFEFEF}123                                                & \cellcolor[HTML]{EFEFEF}117                                                  & \cellcolor[HTML]{EFEFEF}14  & \cellcolor[HTML]{EFEFEF}130                                               & \cellcolor[HTML]{EFEFEF}132                                                   & \cellcolor[HTML]{EFEFEF}71~\%                                                      \\
                                                                                 & \textbf{Effect 1}                             & 1946                         & 22814                                                                      & 22115                                                                        & 1556                         & 18370                                                                      & 17832                                                                         & 194                         & 2210                                                                       & 2125                                                                         & 196                         & 2234                                                                      & 2158                                                                          & 90~\%                                                                               \\
                                                                                 & \cellcolor[HTML]{EFEFEF}\textbf{Effect 2}     & \cellcolor[HTML]{EFEFEF}614  & \cellcolor[HTML]{EFEFEF}5384                                               & \cellcolor[HTML]{EFEFEF}5426                                                 & \cellcolor[HTML]{EFEFEF}483  & \cellcolor[HTML]{EFEFEF}4169                                               & \cellcolor[HTML]{EFEFEF}4200                                                  & \cellcolor[HTML]{EFEFEF}65  & \cellcolor[HTML]{EFEFEF}573                                                & \cellcolor[HTML]{EFEFEF}578                                                  & \cellcolor[HTML]{EFEFEF}66  & \cellcolor[HTML]{EFEFEF}642                                               & \cellcolor[HTML]{EFEFEF}648                                                   & \cellcolor[HTML]{EFEFEF}81~\%                                                      \\
                                                                                 & \textbf{Effect 3}                             & 138                          & 1129                                                                       & 1142                                                                         & 113                          & 952                                                                        & 958                                                                           & 13                          & 80                                                                         & 82                                                                           & 12                          & 97                                                                        & 102                                                                           & 78~\%                                                                              \\
                                                                                 & \cellcolor[HTML]{EFEFEF}\textbf{Not Relevant} & \cellcolor[HTML]{EFEFEF}664  & \cellcolor[HTML]{EFEFEF}6667                                               & \cellcolor[HTML]{EFEFEF}6371                                                 & \cellcolor[HTML]{EFEFEF}537  & \cellcolor[HTML]{EFEFEF}5407                                               & \cellcolor[HTML]{EFEFEF}5175                                                  & \cellcolor[HTML]{EFEFEF}60  & \cellcolor[HTML]{EFEFEF}631                                                & \cellcolor[HTML]{EFEFEF}595                                                  & \cellcolor[HTML]{EFEFEF}67  & \cellcolor[HTML]{EFEFEF}629                                               & \cellcolor[HTML]{EFEFEF}601                                                   & \cellcolor[HTML]{EFEFEF}74~\%                                                      \\
                                                                                 & \cellcolor[HTML]{FFFFFF}\textbf{And}           & \cellcolor[HTML]{FFFFFF}744  & \cellcolor[HTML]{FFFFFF}2799                                               & \cellcolor[HTML]{FFFFFF}2807                                                 & \cellcolor[HTML]{FFFFFF}590  & \cellcolor[HTML]{FFFFFF}2215                                               & \cellcolor[HTML]{FFFFFF}2221                                                  & \cellcolor[HTML]{FFFFFF}80  & \cellcolor[HTML]{FFFFFF}297                                                & \cellcolor[HTML]{FFFFFF}298                                                  & \cellcolor[HTML]{FFFFFF}74  & \cellcolor[HTML]{FFFFFF}287                                               & \cellcolor[HTML]{FFFFFF}288                                                   & 93~\%                                                                              \\
\multirow{-9}{*}{\textbf{\begin{tabular}[c]{@{}c@{}}Top \\ Layer\end{tabular}}}  & \cellcolor[HTML]{EFEFEF}\textbf{Or}            & \cellcolor[HTML]{EFEFEF}230  & \cellcolor[HTML]{EFEFEF}826                                                & \cellcolor[HTML]{EFEFEF}826                                                  & \cellcolor[HTML]{EFEFEF}182  & \cellcolor[HTML]{EFEFEF}667                                                & \cellcolor[HTML]{EFEFEF}667                                                   & \cellcolor[HTML]{EFEFEF}26  & \cellcolor[HTML]{EFEFEF}87                                                 & \cellcolor[HTML]{EFEFEF}87                                                   & \cellcolor[HTML]{EFEFEF}22  & \cellcolor[HTML]{EFEFEF}72                                                & \cellcolor[HTML]{EFEFEF}72                                                    & \cellcolor[HTML]{EFEFEF}91~\%                                                      \\ \midrule
                                                                                 & \textbf{Variable}                              & 1946                         & 26076                                                                      & 25753                                                                        & 1556                         & 20896                                                                      & 20599                                                                         & 194                         & 2543                                                                       & 2509                                                                         & 196                         & 2637                                                                      & 2645                                                                          & 87~\%                                                                              \\
                                                                                 & \cellcolor[HTML]{EFEFEF}\textbf{Condition}     & \cellcolor[HTML]{EFEFEF}1946 & \cellcolor[HTML]{EFEFEF}34927                                              & \cellcolor[HTML]{EFEFEF}34974                                                & \cellcolor[HTML]{EFEFEF}1556 & \cellcolor[HTML]{EFEFEF}27653                                              & \cellcolor[HTML]{EFEFEF}27738                                                 & \cellcolor[HTML]{EFEFEF}194 & \cellcolor[HTML]{EFEFEF}3513                                               & \cellcolor[HTML]{EFEFEF}3498                                                 & \cellcolor[HTML]{EFEFEF}196 & \cellcolor[HTML]{EFEFEF}3761                                              & \cellcolor[HTML]{EFEFEF}3738                                                  & \cellcolor[HTML]{EFEFEF}81~\%                                                      \\
\multirow{-3}{*}{\textbf{\begin{tabular}[c]{@{}c@{}}Lower\\ Layer\end{tabular}}} & \textbf{Negation}                              & 363                          & 1458                                                                       & 1513                                                                         & 287                          & 1154                                                                       & 1199                                                                          & 34                          & 133                                                                        & 139                                                                          & 42                          & 171                                                                       & 175                                                                           & 90~\%                                                                               \\ \bottomrule
\end{tabular}}
\end{table*}

\subsection{Extraction of Conditionals}\label{sec:extraction}
\paragraph{Problem} We define the extraction of conditionals as a sequence labeling problem, in which we are given a certain NL sentence $\mathcal{X}$ in the form of a sequence of $n$ tokens $\mathcal{X} = \{x_i\}_{i=1}^n$ and we are required to produce a sequence $\mathcal{Y}$ of corresponding token labels. Specifically, we aim to demarcate tokens that are relevant for test case derivation from tokens that should be excluded from further processing. In our case, we are interested in twelve token labels. Since conditionals in requirements usually consist of up to three causes and effects~\cite{fischbachREFSQ}, we create individual labels for each cause and effect to clearly separate and map them easily to a CEG:

\begin{enumerate}
        \item \textbf{Cause 1}
  \item \textbf{Cause 2} \smash{{$\left.\rule{0pt}{.5\dimexpr3\baselineskip+2\itemsep+2\parskip}\right\}\text{Cause Labels}$}}
  \item \textbf{Cause 3} 
          \item \textbf{Effect 1}
  \item \textbf{Effect 2} \smash{{$\left.\rule{0pt}{.5\dimexpr3\baselineskip+2\itemsep+2\parskip}\right\}\text{Effect Labels}$}}
  \item \textbf{Effect 3} 
    \item \textbf{Not Relevant} Marks parts of a conditional that are not relevant for automatic test derivation.
    \item \textbf{And} Marks a conjunctive link between two adjacent causes or effects.
    \item \textbf{Or} Marks a disjunctive link between two adjacent causes or effects.\looseness=-1
    \item \textbf{Variable} Marks the variable of a cause or effect.
    \item \textbf{Condition} Marks the condition of a cause or effect.
    \item \textbf{Negation}: Marks negated causes or effects.
    \end{enumerate}  
    
We use these token labels to generate two annotation layers (see \autoref{approach_overview}). The top layer represents the composition of the sentence by specifying the causes, effects, and their combinatorics based on the labels 1 - 9. At the lower layer, we use the labels 10 - 12 to annotate the causes and effects more fine-grained. Consequently, we assign at least one label and at most two labels to a token.

\paragraph{Novelty} 
Contrary to the first part of \textsf{CiRA} (the detection of conditionals) we do not build on previous work for the fine-grained extraction of conditionals and present a new approach in this paper. 

\subsubsection{\textbf{Corpus Creation}}

\paragraph{Data Collection} To train our extraction approach, we require an annotated data set, in which the combinatorics of causes and effects as well as their variables and conditions are labeled. Existing data sets~\cite{xu-etal-2020-review} are not suitable for our use case: The SemEval-2007~\cite{semeval07} and SemEval-2010~\cite{semeval10} data sets contain only single word causal pairs. In the data set presented by Dasgupta et al.~\cite{dasgupta-etal-2018-automatic-extraction}, causes and effects are only coarsely annotated (i.e., connectives, variables, and conditions are not labeled). Due to the unavailability of adequate data, we create our own training corpus. To this end, we build on the data set that we have already used for training our detection algorithm by randomly selecting a subset of the causal requirements (1,946) and annotating them using our twelve predefined labels.

\paragraph{Annotation Process} We involve four annotators with previous experience in the interpretation of conditionals and conduct a workshop where we discuss several examples. To ensure consistent annotations, we create an annotation guideline, in which we define each label along with a set of sample annotations. We use the web-based \textit{brat} annotation platform~\cite{brat12} for labeling each sentence.

\paragraph{Annotation Validity}
To verify the reliability of the annotations, we calculate the inter-annotator agreement. We distribute the 1,946 causal sentences among four annotators, ensuring that 390 sentences are labeled by two annotators (overlapping quote of $\approx$20~\%). Similar to other studies~\cite{kolditz19} that also utilize \textit{brat} to annotate sentences, we calculate the pair-wise averaged $F_{1}$ score~\cite{hripcsak05} based on the overlapping sentences. Specifically, we treat one rater as the subject and the other rater's answers as if they were a gold standard. This allows us to calculate Precision and Recall for their annotations. We then determine the $F_{1}$ score as the harmonic mean of Recall and Precision. We calculate the $F_{1}$ score pairwise between all raters. Subsequently, we take the average of $F_{1}$ scores among all pairs of raters to quantify the agreement of our raters: The higher the average $F_{1}$ score, the more the raters agree with each other. For most of our labels, we obtain an inter-annotator agreement of at least 81~\% (see \autoref{tab:dataanalysis}). The lowest agreement is achieved for \textbf{Cause 2} and \textbf{Cause 3} ($F_{1}$ score of 71~\%). The annotators do not always agree on how granular some expressions should be labeled (e.g., does a text fragment represent another cause, or is it still part of the previous cause?). The highest agreement is measured for the assignment of \textbf{And} ($F_{1}$ score of 93~\%). Averaged across all labels, we achieve an $F_{1}$ score of 83~\%. Based on the achieved inter-annotator agreement values, we assess our labeled data set as reliable and suitable for the implementation of our conditional extraction approach.

\paragraph{Data Analysis}
\autoref{tab:dataanalysis} shows that the majority of our sentences contain only a single cause and effect. About one-third of the sentences contain more complex conditionals comprising two causes or two effects. Only a few sentences contain three causes or three effects. We found that causes and effects are more often connected by a conjunction than by a disjunction. Negated causes and effects occur in about 18~\% of the sentences. At the token level, expressions labeled as \textbf{Effect 1} are often longer than \textbf{Cause 1} expressions. We observe a similar trend at the lower layer. \textbf{Conditions} are usually longer than the \textbf{Variables} of the causes and effects. Across all classes, our data set is strongly unbalanced with four minority classes: \textbf{Cause 3}, \textbf{Effect 3}, \textbf{Or}, and \textbf{Negations}. 

\subsubsection{\textbf{Solution}}\label{solutionExtraction}
Our sequence labeling problem can be solved in two ways: One way is to consider the creation of both annotation layers as two separate multi-class classification tasks. Accordingly, we train two models, where the first model is responsible for recognizing causes and effects, while the second model splits the causes and effects into variables and conditions. In this case, the first model produces the top layer and the second model creates the lower layer. Alternatively, we treat our annotation task as a multi-label classification problem. Consequently, we train only one model, which considers all labels during the prediction and assigns multiple labels to a token. The difference between our multi-class and multi-label solution can be illustrated using the token \enquote{A} shown in \autoref{approach_overview}. In the case of our multi-class solution, the first model assigns the label \textbf{Cause 1} to the token, while the second model assigns the label \textbf{Variable} to the token. Conversely, in the case of our multi-label solution, both labels are assigned by a single model.

\begin{table}
\caption{Overview of architecture and evaluation results of all trained models. Best \textit{macro}-\bm{$F_{1}$} score is marked in \textbf{bold}. Tuned hyperparameters are reported in terms of batch size (bs), learning rate (lr), dropout (d), and size of hidden state.}\label{tab:overallcomparison}
\resizebox{\columnwidth}{!}{%
\begin{tabular}{@{}cllclcl@{}}
\toprule
\multicolumn{1}{l}{}                   & \textbf{}                 & \multicolumn{3}{c}{\textbf{Architecture}}                                                       & \multicolumn{1}{c}{\textbf{}} & \textbf{}                                                                                                                                                                                   \\ \cmidrule(lr){3-5}
\multicolumn{1}{l}{}                   & \keys{model \textbf{\#}}               & Embedding                          & BiLSTM             & Inference                       & \textit{Macro}-$F_{1}$                      & Optimal Hyperparameters                                                                                                                                                                     \\ \midrule
                                       & \cellcolor[HTML]{EFEFEF}\keys{model \Romannum{1}} & \cellcolor[HTML]{EFEFEF}BERT       & \cellcolor[HTML]{EFEFEF}\ding{56} & \cellcolor[HTML]{EFEFEF}Softmax & \cellcolor[HTML]{EFEFEF}76~\%  & \cellcolor[HTML]{EFEFEF}\renewcommand*{\arraystretch}{0.4}\begingroup\begin{tabular}[c]{@{}l@{}}top layer model: \\ bs= 32, lr =  7.49e-05, d = 0.27\\ \\ lower layer model: \\ bs = 64, lr =  6.24e-05, d = 0.18\end{tabular}\endgroup \\
                                       & \keys{model \Romannum{2}}                         & RoBERTa                            & \ding{56}                         & Softmax                         & 75~\%                          & \renewcommand*{\arraystretch}{0.4}\begingroup\begin{tabular}[c]{@{}l@{}}top layer model:\\ bs = 32, lr =  6.24e-05, d = 0.33\\ \\ lower layer model:\\ bs = 32, lr =  4.28e-05, d = 0.21\end{tabular}\endgroup                          \\
\multirow{-3}{*}[6em]{\begin{sideways}\textbf{2x Multi-class Models}\end{sideways}} & \cellcolor[HTML]{EFEFEF}\keys{model \Romannum{3}} & \cellcolor[HTML]{EFEFEF}DistilBERT & \cellcolor[HTML]{EFEFEF}\ding{56} & \cellcolor[HTML]{EFEFEF}Softmax & \cellcolor[HTML]{EFEFEF}78~\%  & \cellcolor[HTML]{EFEFEF}\renewcommand*{\arraystretch}{0.4}\begingroup\begin{tabular}[c]{@{}l@{}}top layer model:\\ bs = 32, lr =  8.80e-05, d = 0.32\\ \\ lower layer model:\\ bs = 64, lr =  9.76e-05, d = 0.36\end{tabular}\endgroup \\ \midrule
                                       & \keys{model \Romannum{4}}                         & BERT                               & \ding{56}                          & Sigmoid                         & 85~\%                          & bs = 64, lr =  8.79e-05, d = 0.26                                                                                                                                                           \\
                                       & \cellcolor[HTML]{EFEFEF}\keys{model \Romannum{5}} & \cellcolor[HTML]{EFEFEF}RoBERTa    & \cellcolor[HTML]{EFEFEF}\ding{56} & \cellcolor[HTML]{EFEFEF}Sigmoid & \cellcolor[HTML]{EFEFEF}\textbf{86~\%}  & \cellcolor[HTML]{EFEFEF}bs = 32, lr = 6.13e-05, d = 0.13                                                                                                                                   \\
                                       & \keys{model \Romannum{6}}                         & DistilBERT                         &  \ding{56}                         & Sigmoid                         & 82~\%                          & bs = 32, lr = 4.47e-05, d = 0.14                                                                                                                                                           \\
                                       & \cellcolor[HTML]{EFEFEF}\keys{model \Romannum{7}} & \cellcolor[HTML]{EFEFEF}BERT       & \cellcolor[HTML]{EFEFEF}\ding{51} & \cellcolor[HTML]{EFEFEF}Sigmoid & \cellcolor[HTML]{EFEFEF}83~\%  & \cellcolor[HTML]{EFEFEF}\begin{tabular}[c]{@{}l@{}}bs = 32, lr = 6.27e-05, \\ lstm\_hidden = 128, d= 0.31\end{tabular}                                                                \\
                                       & \keys{model \Romannum{8}}                         & RoBERTa                            &   \ding{51}                       & Sigmoid                         & 84~\%                          & \begin{tabular}[c]{@{}l@{}}bs = 32, lr = 4.34e-05, \\ lstm\_hidden = 128, d = 0.01\end{tabular}                                                                                      \\
\multirow{-6}{*}[2.5em]{\begin{sideways}\textbf{1x Multi-label Model}\end{sideways}} & \cellcolor[HTML]{EFEFEF}\keys{model \Romannum{9}} & \cellcolor[HTML]{EFEFEF}DistilBERT & \cellcolor[HTML]{EFEFEF}\ding{51} & \cellcolor[HTML]{EFEFEF}Sigmoid & \cellcolor[HTML]{EFEFEF}72~\%  & \cellcolor[HTML]{EFEFEF}\begin{tabular}[c]{@{}l@{}}bs = 32, lr = 9.437e-05, \\ lstm\_hidden = 128, d = 0.50\end{tabular}                                                              \\ \bottomrule
\end{tabular}}
\end{table}    

\paragraph{Models}
In our experiments, we compare the performance of nine different models for multi-class and multi-label classification. \autoref{tab:overallcomparison} provides an overview of their architectures consisting of three different layers: 1) embedding layer, 2) BiLSTM layer, and 3) inference layer. 

\textbf{1) Embedding Layer} Similar to the detection of conditionals, we also use contextual word embeddings for their extraction. However, we do not only perform experiments with BERT but also investigate the influence of RoBERTa~\cite{roberta} and DistilBERT~\cite{distillbert} embeddings on the performance of our models. DistilBERT represents the distilled version of BERT that allows for faster training. RoBERTa is a tuned version of BERT, which shows better prediction performance on various benchmarks but negatively affects training and inference time. To extract the conditionals in fine-grained form, we need to predict on the token level. Hence, in contrast to the detection algorithm where we consider only the CLS token during classification, we pass each token to the classifier, which assigns a label to each token. We consider both actual tokens of a sentence and synthetically added tokens (PAD, SEP, and CLS) because initial experiments demonstrated that the exclusion of synthetic tokens results in significant performance degradation (loss of $\approx$5~\% in \textit{macro}-$F_{1}$). We hypothesize that the performance degradation stems from the fact that the synthetic tokens contain valuable information about the syntax and semantics of a sentence helping the model to comprehend the meaning of a sentence. The number of tokens per class differs depending on the applied tokenizer (see \autoref{tab:dataanalysis}). BERT and DistilBERT use WordPiece as a subword tokenization algorithm, while RoBERTA employs byte-pair encoding (BPE). Nevertheless, both tokenizers differ only slightly, so we set the same maximum length of tokens per sentence for all models. By analyzing the annotated conditionals, a maximum length of 80 tokens proved to be reasonable.

\textbf{2) Bidirectional-LSTM Layer} For \keys{models \Romannum{7} - \Romannum{9}}, we feed the word vectors into a Bidirectional LSTM (BiLSTM) to obtain a hidden state for each word. BiLSTMs have demonstrated to be well suited for sequence labeling problems, because they consider both the past and future contexts of the words. To enable the hidden states to capture both historical and future context information, we train two LSTMs on the input sequence. The forward LSTM processes the sentence from $v_1$ to $v_n$, while a backward LSTM processes from $v_n$ to $v_1$. Consequently, we obtain two hidden states at each time step $t$. $\vec{h}$ is computed based on the previous hidden state $\vec{h}_{t-1}$ and the input at the current step $v_t$, while $\cev{h}$ is computed based on the future hidden state $\cev{h}_{t+1}$ and the input at the current step $v_t$. We obtain the final hidden state by concatenating the forward and backward context representations: $h_i = \vec{h_i} \oplus \cev{h_i}$

\textbf{3) Inference Layer}
For \keys{models \Romannum{1} - \Romannum{3}}, we put the word vectors into a single-layer feedforward neural network that outputs the final predicted tag sequence for the input sentence. Specifically, we use a softmax layer, which calculates the class probabilities for each token: $\hat y = \text{softmax}(Wv_i + b)$, where $\hat y$ are the predicted label probabilities for the $i$-th token, $W$ is the weighted matrix, and $b$ is the bias. We select the class with the highest probability as the final classification result. Since we train two different models for the annotation of the top and lower layer, we apply two different softmax functions. The model predicting the top layer considers nine labels (see \autoref{eqn:top}) while the lower layer is annotated with only three labels (see \autoref{eqn:lower}). In essence, the first model aims to identify causes, effects, and their combinatorics by assigning \textbf{Cause 1-3}, \textbf{Effect 1-3}, and \textbf{Or/And} labels. The second model focuses on the decomposition of causes and effects by assigning \textbf{Variable} and \textbf{Condition} labels.

\noindent\begin{minipage}{.5\linewidth}
\begin{myequation}
\label{eqn:top}
 \text{softmax}(x_i) = \frac{exp(x_i)}{\sum_{j=1}^{9} exp(x_j)}
\end{myequation}
\end{minipage}%
\begin{minipage}{.5\linewidth}
\begin{myequation}
\label{eqn:lower}
  \text{softmax}(x_i) = \frac{exp(x_i)}{\sum_{j=1}^{3} exp(x_j)}
\end{myequation}
\end{minipage}
\strut

In case of the \keys{models \Romannum{4} - \Romannum{6}}, we use a sigmoid layer to perform multi-label classification: $\hat y = \text{sigmoid}(Wv_i + b)$.
We select the classes with a probability $ \geq 0.5 $ as the final classification result. In case of the \keys{models \Romannum{7} - \Romannum{9}}, we consider the hidden states as the feature vectors. Consequently, we define the sigmoid layer as: $\hat y = \text{sigmoid}(Wh_i + b)$
    
\subsubsection{\textbf{Experiments}}
\paragraph{Evaluation Procedure} We follow the idea of \textit{Cross Validation} and divide the data set (1,946 sentences) into a training (1,556), validation (194), and test (196) set. Each class is equally represented across all three data sets, which helps to avoid bias in the prediction (see \autoref{tab:dataanalysis}). We opt for 10-fold cross-validation as a number of studies have shown that a model that has been trained this way demonstrates low bias and variance~\cite{James13}. We use Precision, Recall, and $F_{1}$ score for evaluating our models. Since a single run of a \textit{k}-fold cross-validation may result in a noisy estimate of model performance, we repeat the cross-validation procedure five times and average the scores from all repetitions.

Our data set is strongly unbalanced. Hence, we need to interpret the evaluation metrics carefully. In particular, it is important to distinguish between \textit{macro} and \textit{micro} averages of the metrics. \textit{Macro}-averaging involves the computation of the metrics per class and then averaging them. Hence, each class is treated equally. \textit{Micro}-averaging combines the contributions of all classes to calculate the mean. Thus, it takes label imbalance into account and favors majority classes. In our use case, all classes are equally important. Predicating a minority class like \textbf{Or} is as crucial as predicting a majority class like \textbf{Cause 1}, because it has a major impact on capturing the combinatorics in a sentence. Therefore, we choose the \textit{macro}-$F_{1}$ score as our main evaluation criterion. 

\paragraph{Hyperparameter Tuning}
The performance of DL models depends heavily on the network architecture, as well as the hyperparameters used. Therefore, we compare the performance of our models using different hyperparameter configurations. To determine the optimal hyperparameters for our models, we use the \textit{Tree-structured Parzen Estimator} algorithm~\cite{bergstra11}. During the training process, we check the validation \textit{macro}-$F_{1}$ score periodically to keep the model’s checkpoint with the best validation performance. We train our models for 50 epochs on the training data with a patience of 5 epochs. \autoref{tab:overallcomparison} shows the best hyperparameters for each model.

\begin{table}
\caption{Performance of \keys{model \Romannum{5}} per individual label. \textit{Macro}-\bm{$F_{1}$} scores of at least 90~\% are marked in \textbf{bold}.}\label{tab:labelcomparison}
\begin{tabular}{@{}cllll@{}}
\toprule
\multicolumn{1}{l}{\textbf{}}                                                     & \textbf{Label Type}                            & \textbf{Precision}           & \textbf{Recall}              & \textbf{Macro \bm{$F_{1}$}}                     \\ \midrule
                                                                                  & \cellcolor[HTML]{EFEFEF}\textbf{Cause 1}      & \cellcolor[HTML]{EFEFEF}92~\% & \cellcolor[HTML]{EFEFEF}89~\% & \cellcolor[HTML]{EFEFEF}\textbf{91~\%} \\
                                                                                  & \textbf{Cause 2}                              & 83~\%                         & 72~\%                         & 77~\%                                  \\
                                                                                  & \cellcolor[HTML]{EFEFEF}\textbf{Cause 3}      & \cellcolor[HTML]{EFEFEF}76~\% & \cellcolor[HTML]{EFEFEF}88~\% & \cellcolor[HTML]{EFEFEF}82~\%          \\
                                                                                  & \textbf{Effect 1}                             & 90~\%                         & 89~\%                         & \textbf{90~\%}                         \\
                                                                                  & \cellcolor[HTML]{EFEFEF}\textbf{Effect 2}     & \cellcolor[HTML]{EFEFEF}83~\% & \cellcolor[HTML]{EFEFEF}85~\% & \cellcolor[HTML]{EFEFEF}84~\%          \\
                                                                                  & \textbf{Effect 3}                             & 57~\%                         & 76~\%                         & 65~\%                                  \\
                                                                                  & \cellcolor[HTML]{EFEFEF}\textbf{Not Relevant} & \cellcolor[HTML]{EFEFEF}91~\% & \cellcolor[HTML]{EFEFEF}92~\% & \cellcolor[HTML]{EFEFEF}\textbf{91~\%} \\
                                                                                  & \textbf{And}                                   & 94~\%                         & 96~\%                         & \textbf{95~\%}                         \\
\multirow{-9}{*}{\textbf{\begin{tabular}[c]{@{}c@{}}Top \\ Layer\end{tabular}}}   & \cellcolor[HTML]{EFEFEF}\textbf{Or}            & \cellcolor[HTML]{EFEFEF}85~\% & \cellcolor[HTML]{EFEFEF}92~\% & \cellcolor[HTML]{EFEFEF}88~\%          \\ \midrule
                                                                                  & \textbf{Variable}                              & 87~\%                         & 92~\%                         & 89~\%                                  \\
                                                                                  & \cellcolor[HTML]{EFEFEF}\textbf{Condition}     & \cellcolor[HTML]{EFEFEF}93~\% & \cellcolor[HTML]{EFEFEF}89~\% & \cellcolor[HTML]{EFEFEF}\textbf{91~\%} \\
\multirow{-3}{*}{\textbf{\begin{tabular}[c]{@{}c@{}}Lower \\ Layer\end{tabular}}} & \textbf{Negation}                              & 79~\%                         & 90~\%                         & 84~\%                                  \\ \bottomrule
\end{tabular}
\end{table}

\paragraph{Results}
We first compare the overall performance of our trained models across all classes. In addition, we study the impact of the different word embeddings and the BiLSTM layer on the model performance. Finally, we investigate the performance of our best model to predict the individual labels.

\textbf{1) Overall Comparison} The achieved \textit{macro}-$F_{1}$ scores demonstrate that all investigated models are able to extract conditional statements in fine-grained form (see \autoref{tab:overallcomparison}). However, we observe significant performance gaps between the multi-class and multi-label models. On average, the multi-class models obtain a \textit{macro}-$F_{1}$ score of 76.34~\% while the multi-label models yield an average \textit{macro}-$F_{1}$ score of 82~\%. Consequently, the multi-label models seem to be more suitable for our use case. \keys{Model \Romannum{5}} demonstrates the best performance with a \textit{macro}-$F_{1}$ score of 86~\%, which represents a performance gain of 8~\% compared to the best multi-class \keys{model \Romannum{3}}. We do not witness a major performance difference among most of the multi-label models. In fact, \keys{model \Romannum{4}} shows a very similar behavior as \keys{model \Romannum{5}} and achieves a \textit{macro}-$F_{1}$ score of 85~\%. \keys{Model \Romannum{9}}, however, represents an outlier and produces the poorest \textit{macro}-$F_{1}$ score of all trained models. 

\textbf{2) Impact of Embeddings} Our experiments reveal that the choice of embeddings has an impact on the prediction performance. The best performance of the multi-class models is achieved by using the DistilBERT embeddings (see \autoref{tab:overallcomparison}). In contrast, the multi-label models show the best performance when building on RoBERTa, regardless of the usage of a BiLSTM. Interestingly, the selection of embeddings has the greatest impact on the models that use a BilSTM for feature extraction. For example, a comparison of the performance of \keys{model \Romannum{8}} and \keys{model \Romannum{9}} reveals a performance gap of 12~\% in \textit{macro}-$F_{1}$. In the case of the other models, the performance differences are considerably smaller: the performance of \keys{model \Romannum{2}} and \keys{model \Romannum{3}} differ by only 3~\% in \textit{macro}-$F_{1}$, while \keys{model \Romannum{5}} and \keys{model \Romannum{6}} deviate by only 4~\% in \textit{macro}-$F_{1}$. 

\textbf{3) Impact of BiLSTM Layer} In our setting, adding the BiLSTM layer did not lead to any performance improvement. In fact, the multi-label models demonstrate better performance without the BiLSTM layer. We hypothesize that our amount of training instances is not adequate to sufficiently train the complex BiLSTM architecture and take advantage of its benefits.

\textbf{4) Label Prediction} \autoref{tab:labelcomparison} indicates that \keys{model \Romannum{5}} is capable of processing both conditional statements consisting of only one cause as well as conditionals with multiple causes. Our model predicts \textbf{Cause 1} with very high Precision and Recall resulting in a \textit{macro}-$F_{1}$ score of 91~\%. For the prediction of \textbf{Cause 2} and \textbf{Cause 3}, our model also performs well by achieving \textit{macro}-$F_{1}$ scores of 77~\% and 82~\%, respectively. Conditionals that contain only one effect or two effects can also be processed well by our model. However, our experiments show that the model lacks certainty in the prediction of \textbf{Effect 3} (\textit{macro}-$F_{1}$ score of only 65~\%). We assume that this stems from its under-representation in our training set. The highest \textit{macro}-$F_{1}$ score is achieved by \keys{model \Romannum{5}} for the prediction of \textbf{And}. Likewise, our model performs well in recognizing tokens representing disjunctions (\textit{macro}-$F_{1}$ score of 88~\%). This indicates that our model is able to understand and extract the combinatorics of causes and effects. In addition, the model performs very well in detecting tokens that are not relevant for test case generation. Our experiments prove that our model performs well in predicting both the top and lower layers. The obtained \textit{macro}-$F_{1}$ scores for \textbf{Variable} and \textbf{Condition} show that our model is able to decompose causes and effects into more granular fragments. In addition, \keys{model \Romannum{5}} reliably identifies negations within the conditional statements. 

\begin{figure}
  \centering
  \includegraphics[trim={0 1.5cm 0 0},width=\linewidth]{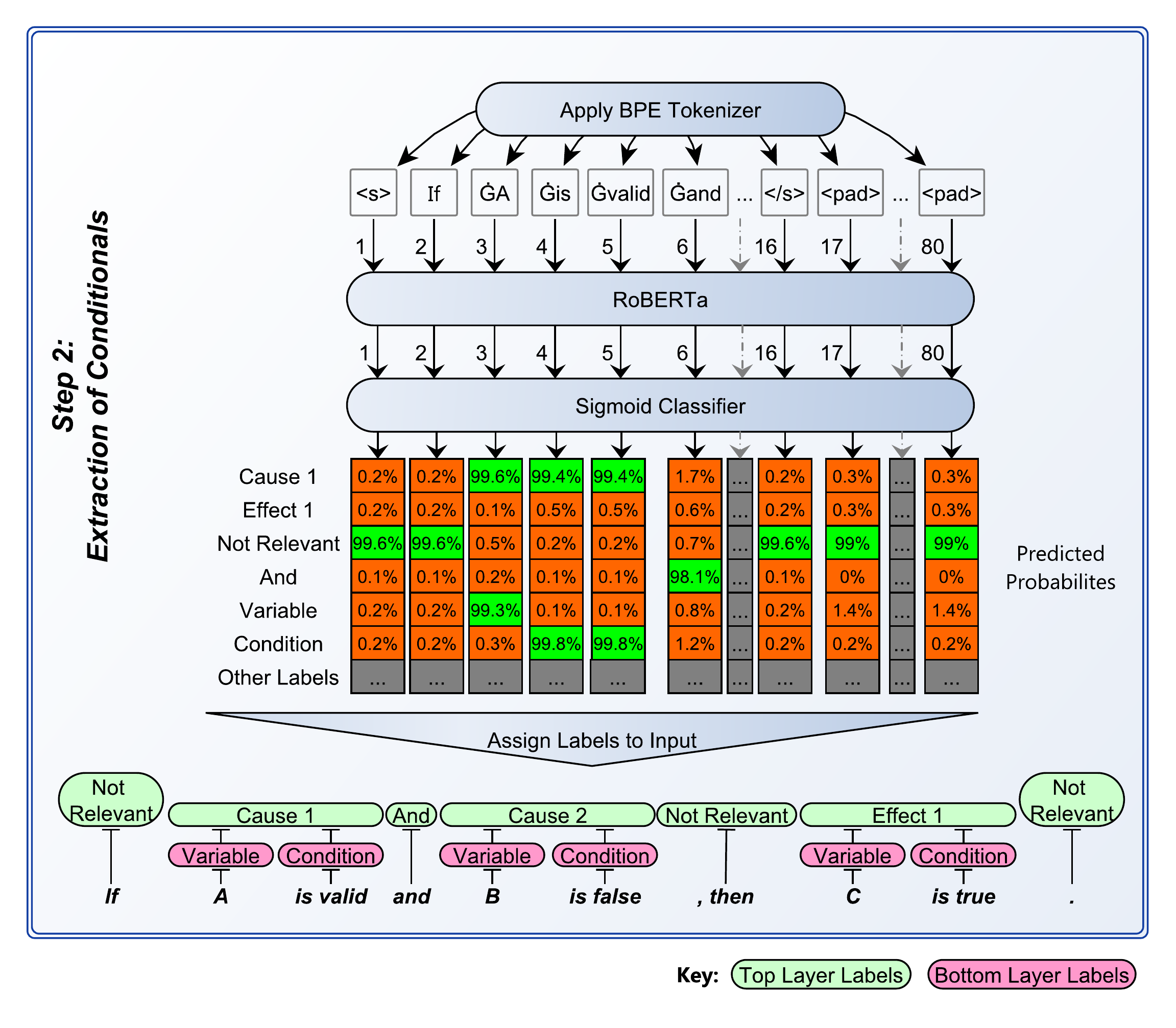}
   \caption{In-depth visualization of the second step in the \textsf{CiRA} pipeline: the fine-grained extraction of conditionals in NL requirements. Processed REQ: \textit{If A is valid and B is false, then C is true.}}
  \label{nurextraction}
\end{figure}

\textbf{5) Summary} Our experiments reveal that \keys{model \Romannum{5}} is best suited to extract conditionals in fine grained form. Specifically, the combination of RoBERTa embeddings (embedding layer) and a sigmoid classifier (inference layer) achieved the best performance. We therefore use \keys{model \Romannum{5}} for the second step in the \textsf{CiRA} pipeline. A detailed insight into the functionality of \keys{model \Romannum{5}} is given in \autoref{nurextraction}.

\paragraph{Running Example}
In the second step, \textsf{CiRA} extracts the conditional statements from the requirements that were classified as causal in the first step. For this purpose, REQ A, REQ B, REQ D, REQ E, REQ F, and REQ H are decomposed into individual tokens using the BPE tokenizer and then converted into RoBERTa embeddings. Subsequently, each token embedding is fed into a sigmoid classifier, which calculates the probability for each of our twelve labels (\textbf{Cause 1}, \textbf{Cause 2}, \dots \textbf{ Condition}) that the token should be associated with that class. We select the classes with a probability $ \geq 0.5 $ as the final classification result. \autoref{fig:runningExampleCira} shows the extracted conditionals by \textsf{CiRA} from the THEMAS requirements. The running example demonstrates that \textsf{CiRA} is able to extract conditionals in fine-grained form - independent of whether the requirements contain simple conditionals consisting of a single cause and effect (see REQ D and REQ F) or complex conditionals with multiple causes and effects (see REQ B). Further, \textsf{CiRA} is able to detect causes and effects in different positions in a sentence, which can be illustrated by REQ E. In this case, \textsf{CiRA} extracts the conditional statement correctly even though \textbf{Cause 1} and \textbf{Cause 2} do not immediately follow each other but instead are separated by \textbf{Effect 1}.

\subsection{Creation of Cause-Effect-Graph}\label{sec:creation}

In the final step, we produce a CEG based on the extracted causes and effects. Creating a CEG is not a trivial task, especially for complex conditional statements consisting of multiple causes and effects. We handle the following cases:
   
        \textbf{\circled{1} Single Cause - Single Effect} In the simplest case, we create two nodes and draw an edge from the cause node to the effect node. 
        
\begin{figure}
  \centering
  \includegraphics[trim={0 1.5cm 0 0},width=\linewidth]{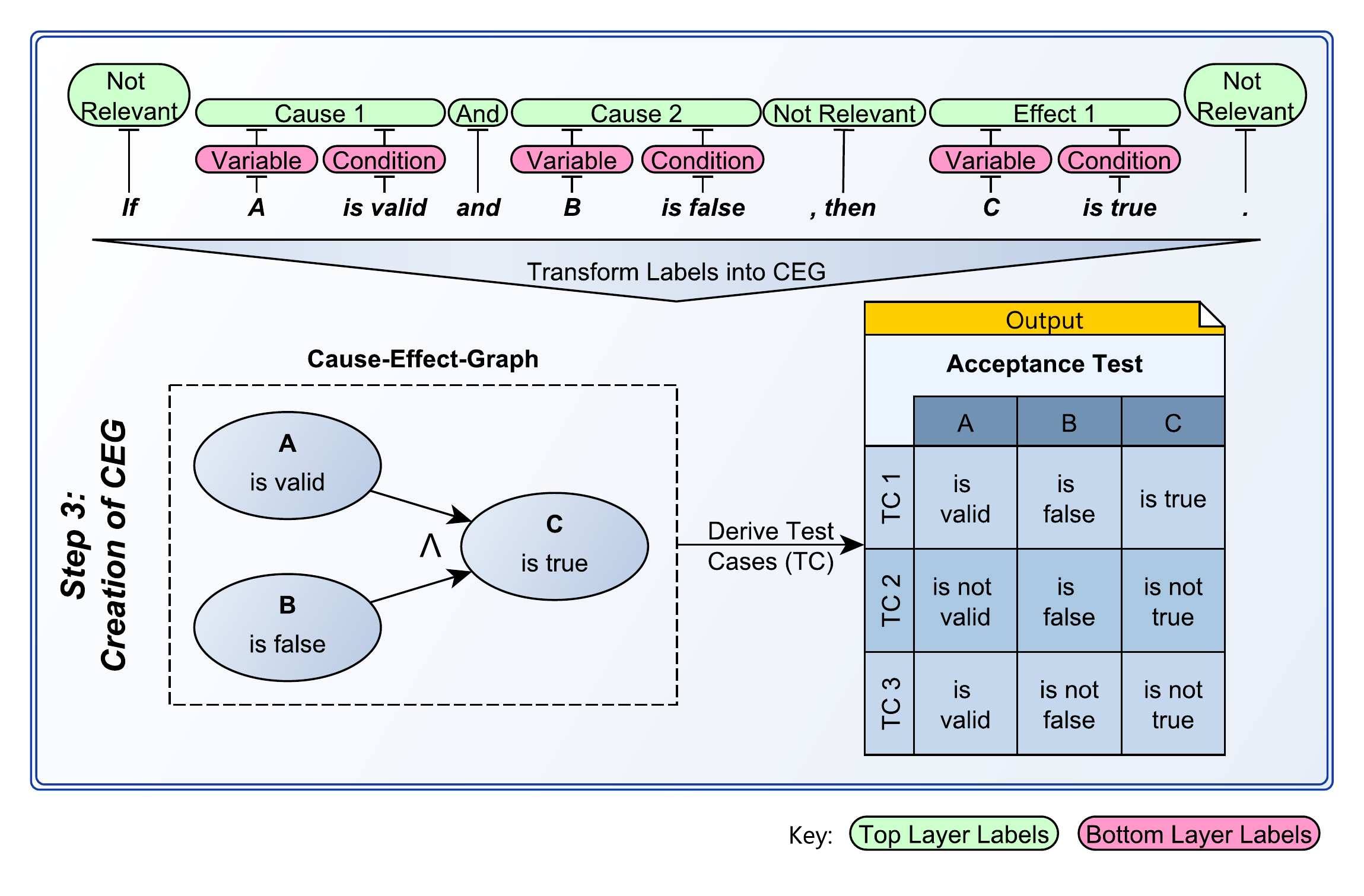}
   \caption{In-depth visualization of the third step in the \textsf{CiRA} pipeline: the creation of a \textit{Cause-Effect-Graph}. Processed REQ: \textit{If A is valid and B is false, then C is true.}}
  \label{nurCEG}
\end{figure}

        \textbf{\circled{2} Multiple conjunctive Causes - Single Effect} In this case, all causes must occur jointly for the effect to occur. Thus, we connect all cause nodes with the effect node using the connective $\wedge$. We illustrate this case by using an exemplary requirement in \autoref{nurCEG}.
        
        \textbf{\circled{3} Multiple disjunctive Causes - Single Effect} In this case, the occurrence of one of the causes is sufficient for the effect to occur. Thus, we link all cause nodes with the effect node using the logical connective $\vee$.
        
        \textbf{\circled{4} Combination of conjunctive and disjunctive Causes - Single Effect} Conditionals are usually not parenthesized, which causes a certain degree of ambiguity when combining conjunctions and disjunctions. To convert such conditionals uniformly into a CEG, we follow the precedence rules of propositional logic. Hence, we evaluate conjunctions with higher precedence than disjunctions. To this end, we create an intermediate node for each set of conjunctive causes and connect the causes with the intermediate node using the logical connective $\wedge$. Subsequently, we connect the disjunctive cause(s) and the intermediate node(s) with the effect using the logical connective $\vee$. If no connective can be found between two adjacent causes, the closest subsequent connective is used. For example, in an enumeration like \enquote{Owners, tenants, and  managers} only the connection between \enquote{tenants} and \enquote{managers} is explicit, whereas \enquote{owners} and \enquote{tenants} are also implicitly connected by a conjunction.
        
       \textbf{\circled{5} Multiple conjunctive Effects} We create a node for each effect and connect them to the causes according to the rules described above. We do not allow effects to be connected with a disjunction as this would denote an indeterministic system behavior.
   
 \textbf{\circled{6} Correction of incomplete nodes} In the simplest case, a cause or effect encompasses both a variable and condition in the lower annotation level. We then fill the created nodes with the corresponding information. If either of the two labels is missing, the information is extracted from the nearest referent instead.
 
\begin{landscape}
\begin{figure}
    \centering
    \includegraphics[width=\linewidth]{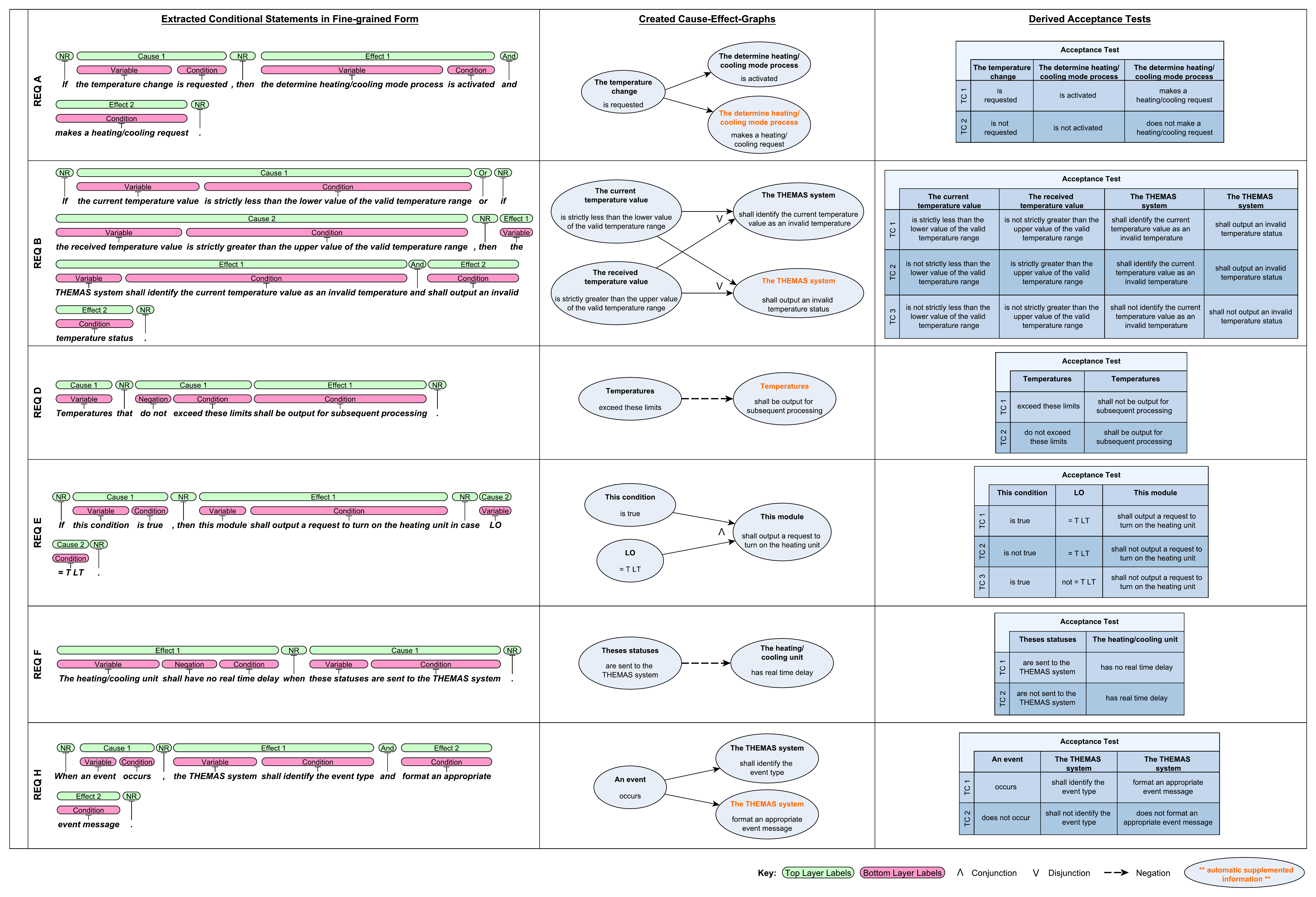}
    \caption{Overview of the conditionals extracted by \textsf{CiRA} in fine grained form (left), the generated \textit{Cause-Effect-Graphs} (center), and the derived acceptance tests (right) per requirement defined in the THEMAS specification.}
    \label{fig:runningExampleCira}
\end{figure}
\end{landscape}

\paragraph{Running Example} 
In the third step, \textsf{CiRA} interprets the conditional statements extracted from REQ A, REQ B, REQ D, REQ E, REQ F, and REQ H, and creates a corresponding CEG. Subsequently, \textsf{CiRA} applies the BPST (cf.~\cite{Myers} and~\cite{Nursimulu95}) to derive the minimum number of required test cases from each CEG. \autoref{fig:runningExampleCira} presents an overview of the CEGs created by \textsf{CiRA} and the respective automatically generated test specifications for each causal requirement included in the THEMAS requirements specification (see \autoref{fig:runningExample}).

The CEG generated for REQ A corresponds to case \circled{5} described above. Specifically, one cause is the trigger for two conjunctive effects: \enquote{the heating/cooling process is activated} and \enquote{[the heating/cooling process] makes a heating/cooling request}. We can observe from the extracted conditional for REQ A that the variable of \textbf{Effect 2} is not explicitly defined in the requirement. We must therefore automatically complete the node of \textbf{Effect 2} by adopting the variable from \textbf{Effect 1} (see case \circled{6}). REQ B specifies complex system behavior and contains two causes and two effects. When creating a corresponding CEG, we need to consider both case \circled{3} and case \circled{5}. Thus, we link all cause nodes with both effect nodes using the logical connective $\vee$. Similar to REQ A, we need to automatically complete the variable of \textbf{Effect 2} since it is not explicitly defined in REQ B (see case \circled{6}). REQ D contains a negated cause that is responsible for the occurrence of a single effect. In other words, not exceeding the limit is required for a temperature to be eligible for further processing. Therefore, we are dealing with case \circled{1} and have to negate the edge between cause and effect. Once again, the variable of an effect node is not described in the requirement. In contrast to REQ A and REQ B we do not complement the node of \textbf{Effect 1} with the variable of a neighboring effect but rather adopt the variable of \textbf{Cause 1}. The CEG generated for REQ E corresponds to case \circled{2}. Hence, we connect both cause nodes with the single effect using the connective $\wedge$. REQ F includes a negated effect that is triggered by a single cause. We thus create the CEG based on the rules described in case \circled{1} and negate the edge between cause and effect. REQ H contains one cause and two conjunctive effects. We create a corresponding CEG by applying the rules described in case \circled{5}. We complement the node of \textbf{Effect 2} by adopting the variable of \textbf{Effect 1}.

\textsf{CiRA} automatically created a total of 14 test cases for all causal requirements included in our running example. The created acceptance tests for REQ D and REQ F are trivial and contain only two parameters that have to be checked. The other acceptance tests are of higher complexity as they contain more input and output parameters: To fully test REQ A, REQ E and REQ H, two test cases including three parameters each have to be checked. The acceptance test created for REQ B involves three test cases with four parameters.

\begin{tcolorbox}[breakable, enhanced jigsaw,arc=0mm,left=1mm,boxsep=1mm,title=\textbf{Answer to RQ 1:}]
\textit{We develop \textsf{CiRA}, an approach to identify and extract conditionals in NL and transfer them into a CEG, from which the minimal set of required test cases can be derived automatically. For the detection, we use syntactically enriched BERT embeddings combined with a softmax classifier~\cite{fischbachREFSQ}. Our experiments show that a sigmoid classifier built on RoBERTA embeddings (see \keys{model \Romannum{5}}) is best suited to extract conditionals in fine-grained form.}
\end{tcolorbox}  

\section{Case Study: \textsf{CiRA} in practice}
\label{sec:evaluation}
\paragraph{Objective} To answer RQ 2 and RQ 3, we conduct a case study in an exploratory fashion. We aim to evaluate whether \textsf{CiRA} could either replace or augment the existing manual approach for creating test cases. For our study, we follow the guidelines by Runeson and Höst~\cite{runeson} for conducting case study research.

\subsection{Case Sampling and Study Objects}
We apply purposive case sampling augmented with convenience sampling~\cite{Kitchenham}. Specifically, we approached some of our industry contacts inquiring whether they are interested in exploring the potential of \textsf{CiRA}. We were provided with data from three companies operating in different domains: \textit{Allianz Deutschland AG} (insurance), \textit{Ericsson} (telecommunication), and \textit{Leopold Kostal GmbH \& Co. KG} (automotive). Since the data is subject to non-disclosure agreements, we are unable to share the provided requirements and test cases.

\paragraph{Allianz Data} We analyze 219 acceptance criteria describing the functionality of a business information system used for vehicle insurance. 127 of these acceptance criteria follow a causal pattern and are therefore suitable for assessing \textsf{CiRA}. The remaining acceptance criteria specify the expected functionality based on process flows (16 criteria) or in a static way (76 criteria). We analyze the acceptance tests that were manually created for each of the causal acceptance criteria. In total, 309 test cases were designed, which corresponds to about 2.43 test cases per acceptance test.

\paragraph{Ericsson Data} We analyze 109 requirements derived from five \textit{Business Use Cases} (BUCs), which are feature-level units of development at Ericsson. The BUCs originate from different functional topics. 49 of these 109 requirements contain conditionals while the remaining requirements are expressed in a static way. In total, 65 test cases were manually generated for the 49 causal requirements, which corresponds to about 1.33 test cases per acceptance test.

\paragraph{Kostal Data} We analyze a requirements specification describing a plug interlock function, which prevents a charging plug from being disconnected during an active charging process of an electric car. The specification includes 135 functional requirements. 79 of these functional requirements are indeed causal while 56 requirements describe the functional behavior in a static way: \enquote{The signal \texttt{signalName} shall be set to \texttt{InitValue}}. In our case study, we focus only on the acceptance tests that were manually created for the 79 causal requirements. In total, 204 test cases were designed, which corresponds to about 2.58 test cases per acceptance test.

\subsection{Study Design}
\paragraph{Approach for RQ 2} We want to study whether \textsf{CiRA} can achieve the status quo or even lead to an improvement of the manual test case derivation. To this end, we pass all study objects through our pipeline and compare the automatically created acceptance tests with the manually created acceptance tests. We assess two acceptance tests to be equal if they contain the same test cases. Two test cases are equivalent if they consist of the same input and output parameters with semantically identical variables and conditions. However, we allow syntactical differences between the test cases (e.g., different spelling of parameters), since they still test the same functionality. By comparing the test cases created by \textsf{CiRA} with the manually created test cases, we found that it is sometimes not possible to establish a one-to-one relationship. Partly, test designers aggregate related parameters, so that a manual test case may cover multiple automated test cases (one-to-many relationship). Therefore, two acceptance tests may also be equivalent even if the number of test cases differs. If we observe discrepancies between a manual acceptance test and an automatic acceptance test, we involve test designers from our case companies and examine the set differences: 1) test cases created exclusively by the manual approach (MA), and 2) test cases generated exclusively by our automated approach (AA). In both cases, we ask the test designers whether a certain test case is required to fully check the functionality described by the requirement to assess its \textit{relevance} (rel). Consequently, we investigate five different categories of test cases:

\begin{itemize}[]
  \item \keys{$Identical$} : A test case that has been created manually as well as automatically by \textsf{CiRA}.
  \item \keys{$ AA \wedge rel$} : A test case that has been missed in manual test design and should be included in the acceptance test.
  \item \keys{$ AA \wedge \neg rel$} : A superfluous test case that is correctly not included in the manually created acceptance test.
  \item \keys{$ MA \wedge rel$} : A test case that has been missed by \textsf{CiRA} and should be included in the acceptance test.
  \item \keys{$ MA \wedge \neg rel$} : A superfluous test case that is correctly not included in the automatically created acceptance test.
\end{itemize}

\paragraph{Approach for RQ 3} To answer the third RQ, we document all errors and ambiguities of \textsf{CiRA} produced during the approach for RQ2 to discuss these with test designers of our case companies. To avoid interviewer bias, we do not involve the test designers who created the respective acceptance tests. Instead, we interview one or more colleagues who are also familiar with the functionalities described in the requirements. To determine the reasons for deviating acceptance tests, we examine the manually and automatically created test case as well as the corresponding requirements jointly with the test designers: a meeting between one of the two first authors and each respective company is scheduled during which all recorded deviations are presented, jointly discussed and their final categorization according to the aforementioned five categories is confirmed. The first author involves two test designers at Allianz and two test designers at Kostal, the second author involves one test designer at Ericsson.

\begin{figure}
  \centering
  \includegraphics[width=\linewidth,trim=15 1cm 10 0]{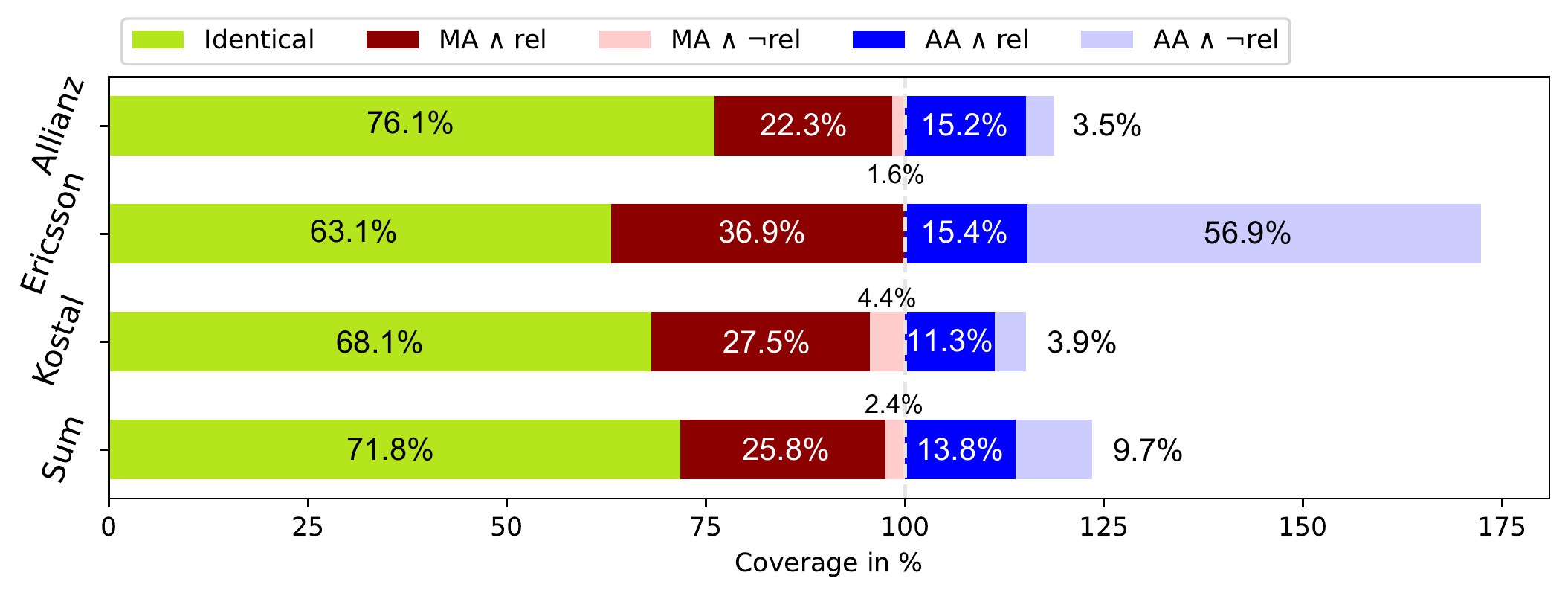}
  \caption{Case study results. Comparison of manually and automatically created test cases.}
  \label{evaluation_results}
\end{figure}

\subsection{Case Analysis}
We report on our findings structured by our research questions (RQ 2 and RQ 3). \autoref{evaluation_results} provides an overview of manually and automatically created test cases.

\subsubsection*{RQ 2: Can \textsf{CiRA} create the same test cases as the manual approach?} 

\paragraph{Findings at Allianz} \textsf{CiRA} detected 90.55~\% of the causal acceptance criteria. Consequently, no test cases were created for the missed 12 causal criteria. For the correctly classified criteria, \textsf{CiRA} generated 314 test cases. This corresponds to about 2.73 test cases per acceptance test. We were able to draw a one-to-one relationship between 224 manually and automatically created test cases. Additionally, we observed a one-to-many relationship between eleven manually created test cases and 32 automatically created test cases. Thus, 76.05~\% of the manually created test cases could be automatically generated. However, 74 test cases were not created by \textsf{CiRA}, of which 27 test cases are related to criteria that were incorrectly identified as non-causal. According to the test designers, the remaining 47 MA test cases can be classified as follows: 42 are necessary to fully test the system functionality while five test cases are superfluous. A comparison of the automatically created test cases with the manually created test cases highlights that 58 test cases have not yet been considered in the manual test design. According to the test designer, these 58 AA test cases can be clustered as follows: 47 are indeed \textit{relevant} while eleven should not be included in the acceptance test.

\paragraph{Findings at Ericsson} \textsf{CiRA} correctly classified 79.6~\% of the causal requirements but failed to do so for ten causal requirements. 91 test cases were automatically generated based on these identified requirements, which corresponds to about 2.33 test cases per acceptance test. 28 manual test cases were automatically created by \textsf{CiRA} in a one-to-one, 13 more in a one-to-many relationship, resulting in an automatic generation of 41 of 65 test cases (63.1~\%). However, 24 test cases were not created by \textsf{CiRA}, of which seven test cases are related to criteria that were incorrectly identified as non-causal. According to the test designer, the remaining 17 MA test cases are all necessary to fully test the system's functionality. A comparison of the automatically created test cases with the manually created test cases highlights that 47 test cases have not yet been considered in the manual test design. According to the test designer, these 47 AA test cases can be clustered as follows: ten are indeed \textit{relevant} while 37 should not be included in the acceptance test.

\paragraph{Findings at Kostal} \textsf{CiRA} correctly classified 72 requirements as causal. However, it failed to identify the remaining seven causal requirements. Hence, no test cases were ultimately created for these requirements. In the case of the correctly classified requirements, \textsf{CiRA} produced 194 test cases. This corresponds to about 2.69 test cases per acceptance test. We found a one-to-one relationship between 122 manually and automatically created test cases. In addition, we were able to draw a one-to-many relationship between 17 manual test cases and 41 automatically created test cases. Thus, 68.14~\% of the manually created test cases could be created automatically. Nevertheless, 65 manually created test cases are not included in the set of automated test cases. 16 of these exclusively manually created test cases refer to the causal requirements that \textsf{CiRA} missed. In the case of the other 49 test cases, we ask test designers at Kostal about their relevance. In fact, 81.63~\% of the exclusively manually created test cases are deemed \textit{relevant}. According to the test designers, nine test cases are superfluous and can be removed from the test set. Examining the automatically created test cases, we observe that 31 test cases have not been considered in the manual creation so far. Interestingly, the test designers confirmed that 74.19~\% of these test cases were indeed missed in the manual process. However, eight exclusively automatically created test cases are not \textit{relevant} and thus are correctly not included in the manual set.

\begin{tcolorbox}[breakable, enhanced jigsaw,arc=0mm,left=1mm,boxsep=1mm,title=\textbf{Answer to RQ 2:}]
\textit{Across all case companies, \textsf{CiRA} automatically created 71.8~\% of the 578 manually created test cases. \textsf{CiRA} was further able to identify 136 test cases that were missed in manual test design. In fact, 58.8~\% of these exclusively automatically generated test cases are indeed relevant and should be included in the acceptance test. We conclude that \textsf{CiRA} is able to automatically create a significant amount of relevant (known and new) test cases.}
\end{tcolorbox}  

\subsubsection*{RQ 3: What are the reasons for deviating test cases?} 

\paragraph{Incomplete requirements} We found that the main reason for test cases that could not be created automatically lies in the poor information available in the requirements. The interviewed test designers confirmed that domain knowledge is often required to determine all relevant test cases. In the case of Kostal, 19 out of 79 requirements were incomplete. We found that \textsf{CiRA} could not generate 37 \keys{$ MA \wedge rel$} test cases due to a lack of information in these requirements. At Allianz, 16 out of 127 causal acceptance criteria lack information. Our analysis shows that \textsf{CiRA} could not generate 31 \keys{$ MA \wedge rel$} test cases due to incomplete acceptance criteria. At Ericsson, 17 \keys{$ MA \wedge rel$} test cases could not be generated due to underspecified or missing requirements. 

\paragraph{Incorrect combinatorics} We noticed that some of the exclusively manually created test cases are superfluous - they can be merged or are already covered by other test cases. The interviews revealed that in these cases the combinatorics of the input and output parameters were interpreted incorrectly. According to the test designers, this stems mainly from the fact that test cases are often not created systematically, but rather based on past experience. Unsystematic test design may not only result in superfluous test cases but can also lead to necessary test cases being ignored. We observed that test designers tend to create positive cases and neglect negative cases. At Kostal, 21 of the 23 \keys{$ AA \wedge rel$} test cases were actually negative cases. Only two positive cases were overlooked in the manual process. At Allianz, 36 of the 47 \keys{$ AA \wedge rel$} test cases were actually negative cases. 11 positive cases were missed by the test designers. In the case of Ericsson, all ten \keys{$ AA \wedge rel$} test cases were overlooked negative test cases.

\paragraph{Infeasible test cases} Our analysis shows that some of the exclusively automatically created test cases can not occur in practice. According to the test designers, this problem arises mainly for negative test cases where certain scenarios are tested that can only occur theoretically. For example, some parameters can not take the value false at the same time, even if this case should be checked from a combinatorial point of view. In the case of Kostal, we found that three of the eight \keys{$ AA \wedge \neg rel$} test cases can not be checked in practice. At Allianz, five of the eleven \keys{$ AA \wedge \neg rel$} test cases can only occur theoretically. At Ericsson, 28 of 37 \keys{$ AA \wedge \neg rel$} test cases fell into this category.

\paragraph{\textsf{CiRA} errors} \textsf{CiRA} produced not only errors in the detection of the conditionals, but also failed in some cases to extract and translate them into the CEG. At Kostal, \textsf{CiRA} failed to generate 3 \keys{$ MA \wedge rel$} test cases and instead created five \keys{$ AA \wedge \neg rel$} test cases, because the generated CEG reflected a wrong causal relation. In the case of Allianz, \textsf{CiRA} failed to create eleven \keys{$ MA \wedge rel$} test cases and instead generated six \keys{$ AA \wedge \neg rel$} test cases. In the case of Ericsson, \textsf{CiRA} produced nine \keys{$ AA \wedge \neg rel$} test cases due to incorrect interpretation of the causal relation. We found that these errors occurred mainly when the conditionals contained three or more effects. This confirms the findings from our experiment that \textsf{CiRA} struggles in reliably identifying more than two effects (see \autoref{tab:labelcomparison}).

\begin{tcolorbox}[breakable, enhanced jigsaw,arc=0mm,left=1mm,boxsep=1mm,title=\textbf{Answer to RQ 3:}]
\textit{In our setting, we observed four reasons for deviating test cases: incomplete requirements, incorrect combinatorics, infeasible test cases, and \textsf{CiRA} errors. We found that incomplete requirements are the main reason for test cases that could not be created automatically by \textsf{CiRA}.}
\end{tcolorbox}

\section{Discussion}
\label{sec:discussion}
This section discusses our results and summarizes both the potentials as well as limitations of \textsf{CiRA}. Based on our discussion, we deduce key take-aways for practitioners.

\subsection{Potentials, Limitations, and Key Take-aways}
Our case study demonstrates that \textsf{CiRA} is able to support practitioners in deriving relevant test cases from conditionals. Across all industry partners, \textsf{CiRA} automatically generates more than 70~\% of the manually created test cases. However, \textsf{CiRA} does not achieve full automation of acceptance test creation, mainly due to incomplete requirements. \textsf{CiRA} is heavily dependent on the information contained in the requirements and consequently unable to create test cases for which additional domain knowledge is required. Thus, our case study confirms the findings of Mendez et al.~\cite{mendez} that incompleteness is still a major problem in practice and hinders the automatic processing of requirements. 

\begin{tcolorbox}[breakable, enhanced jigsaw,arc=0mm,left=1mm,boxsep=1mm,title=\textbf{1. Key Take-away:}]
\textit{In fact, \textsf{CiRA} can help to generate acceptance tests automatically. However, \textsf{CiRA} does not substitute a test designer since domain knowledge is often necessary to identify all required test cases.}
\end{tcolorbox}

According to the test designers, the main benefit of \textsf{CiRA} is its ability to create test cases automatically based on heuristics. Hence, it is independent of human bias and able to identify test cases that may be missed in the manual process. We argue that \textsf{CiRA} should always be used as a supplement to the existing manual process to highlight all test cases that should be tested from a combinatorial point of view, in particular negative test cases that were proportionally more often overlooked than positive test cases. The automatically generated set of test cases may then be manually extended by test cases that require domain knowledge. At Ericsson, we observed that a large amount of automatically generated test cases were irrelevant since they can only occur theoretically. Hence, when utilizing \textsf{CiRA} as a supplement to manual test design, test designers need to filter the automatically generated test cases. We however argue that this does not impede the usability of \textsf{CiRA}, as it is significantly easier to manually discard irrelevant test cases than to manually identify undetected, relevant test cases. We favor Recall over Precision since the performance gained due to the automatization outweighs the manual effort to filter the resulting set of proposed test cases.

\begin{tcolorbox}[breakable, enhanced jigsaw,arc=0mm,left=1mm,boxsep=1mm,title=\textbf{2. Key Take-away:}]
\textit{\textsf{CiRA} is particularly useful for automatically identifying negative test cases, which are often overlooked in the manual creation process. However, not all test cases created by \textsf{CiRA} are necessarily relevant, requiring subsequent manual review of the automatically created test specifications.}
\end{tcolorbox}

Previous approaches~\cite{fischbachICST,Sorgente2013} for extracting conditionals analyze the grammatical structure of a sentence by dependency parsing and extract the conditionals from the dependency tree. However, dependency trees usually contain incorrect arcs for sentences that contain grammar mistakes causing the approaches to extract wrong conditionals or even to fail completely. Since \textsf{CiRA} decomposes each sentence using subword tokenization and labels each token individually, it is much more robust against grammar errors and is also able to process out-of-vocabulary words. Nevertheless, studies~\cite{kumar-etal-2020-noisy} reveal that language models such as BERT show significant performance degradation with increasing amounts of noisy data. As a result, we hypothesize that \textsf{CiRA's} robustness against grammatical mistakes is limited to a few errors in a sentence. We, therefore, propose to combine \textsf{CiRA} with requirements smell checkers~\cite{FEMMER2017190} in the future to automatically verify the linguistic quality of requirements before passing them into the \textsf{CiRA} pipeline. 

\begin{tcolorbox}[breakable, enhanced jigsaw,arc=0mm,left=1mm,boxsep=1mm,title=\textbf{3. Key Take-away:}]
\textit{Fully automated acceptance test generation is difficult to achieve because requirements often suffer from poor quality. RE teams should therefore first check the quality of the requirements before processing them with \textsf{CiRA}.}
\end{tcolorbox}

\textsf{CiRA} is limited to single sentence causality and is not able to extract conditional statements that span multiple sentences. However, two-sentence causality may arise in practice (e.g., indicated by \enquote{therefore}, \enquote{hence}), requiring us to extend \textsf{CiRA} in future work. According to the test designers, a further challenge in the extraction of conditionals relates to the handling of \textit{event chains} (i.e., linked causal requirements, in which the effect of a conditional represents a cause in another conditional). In such cases, it is no longer sufficient to create a single CEG. Rather, we must create several CEGs and connect them to each other. Currently, \textsf{CiRA} only allows the creation of acceptance tests for causal requirements. For full automation of test case design, however, we also require approaches capable of processing static requirements and process flows. 

\begin{tcolorbox}[breakable, enhanced jigsaw,arc=0mm,left=1mm,boxsep=1mm,title=\textbf{4. Key Take-away:}]
\textit{So far, the feasibility of \textsf{CiRA} is limited to single causal requirements. As a consequence, we still need to develop methods for the automatic generation of test cases from static requirements and process flows.}
\end{tcolorbox}

Our case study focuses on a quantitative comparison between manually and automatically created test cases. However, several other metrics are available to benchmark test cases~\cite{tran21}. For example, structural criteria like \textit{test understandability} investigate whether a test is easy to understand in terms of its internal and external descriptions. We plan to extend our study to obtain further insights into the quality of the test cases generated by \textsf{CiRA}.

\subsection{Threats to Validity}

As in every empirical study, our case study (see RQ 2 and RQ 3) is also subject to potential validity threats. This section discusses these threats and describes how we mitigated them.

\paragraph{Internal Validity} 
We acknowledge a possible threat to internal validity due to selection bias of suitable requirements artifacts. In all cases, the artifact selection was driven by the availability of data. Hence, requirements and test cases were not actively sampled to improve \textsf{CiRA's} performance.

\paragraph{Construct Validity}
The comparison between the manually and automatically created test cases might be subject to researcher bias. To mitigate this risk, the first and second authors individually mapped the test cases. Subsequently, the mapping was cross-checked and discussed within the research group. A further threat to internal validity is the potential bias of the interviewed test designers. To keep this risk as low as possible, we interviewed each test designer independently and compared the reasons for the deviating test cases.

\paragraph{External Validity} To achieve reasonable generalizability, we selected requirements and test cases from different domains. However, the limited sample size does not provide the statistical basis to generalize the results of our study beyond the studied case companies. Nevertheless, we hypothesize that \textsf{CiRA} may also be valuable for other companies considering that conditionals are widely used in requirements. Validation of this claim requires further empirical investigation.

\section{Related Work}

Since the early 1980s, NLP techniques have been applied to RE artifacts to support a variety of use cases: e.g., requirements classification~\cite{Hey20}, topic modeling~\cite{Guelle20}, and quality checks~\cite{FEMMER2017190}. A comprehensive overview of existing NLP4RE tools is provided by Zhao et al.~\cite{Zhao21}. In this paper, we use NLP methods to extract conditionals from requirements in fine-grained form and to derive test cases automatically. This section reviews existing approaches for both use cases and relates them to our work.

\label{sec:related}
\paragraph{Approaches to Test Case Derivation} There is a rich body of work on automatically deriving test cases from semi-formal and formal requirements~\cite{Ahsan17}. Wang et al.~\cite{wangbriand} and Zhang et al.~\cite{Zhang14} describe how to automatically generate executable system test cases from restricted use case models. Sarmiento et al.~\cite{Sarmiento} explain how semi-formal requirements can be translated into Petri-Net models, which can be used as input for test scenario generation. Carvalho et al.~\cite{Carvalho14} present their tool NAT2TEST$_{SCR}$ and show how to automatically generate test cases from requirements written in SysReq-CNL. However, only a few approaches allow to automatically create test cases from informal requirements. Santiago J\'{u}nior and Vijaykumar~\cite{santiagoSpace} present SOLIMVA capable of translating NL requirements into state charts used for the eventual test case generation. Verma and Beg~\cite{Verma} describe a similar approach for translating informal requirements into knowledge representation graphs. Dwarakanath and Sengupta~\cite{Dwarakanath12} present Litmus, a tool that applies a syntactic parser called \textit{Link Grammar} in order to analyze the structure of an NL requirement and create test cases accordingly. Goffi et al.~\cite{Goffi2016} demonstrate how to create test cases for exceptional behaviors from Javadoc comments. The mentioned approaches have three major drawbacks: (1) they require manual work such as the creation of a dictionary~\cite{santiagoSpace} defining the application domain in which the approach will be used, (2) they do not ensure that only the minimal number of required test cases is created, and (3) they lack tool-support and are thus not immediately usable for practitioners. In previous work~\cite{fischbachICST}, we address this research gap and present \textit{Specmate} that automatically converts acceptance criteria into test cases by extracting cause-effect-relations based on dependency parsing. However, we found that \textit{Specmate} is not robust against grammatical errors and fails to process words that are not yet part of its training vocabulary. In this paper, we therefore shift from using dependency parsing to solving the extraction of conditionals as a sequence labeling problem.

\paragraph{Approaches to Extract Conditionals}
Several approaches for conditional extraction have been developed~\cite{yang2021survey}: Rule-based approaches~\cite{khoo98, puente} extract conditionals by applying linguistic patterns such as \textit{[cause] and because of this, [effect]}. Their performance relies on hand-coded patterns, which require extensive manual work. Other approaches use ML. Chang and Choi~\cite{chang05} use a Naive Bayes classifier to predict the probability of a causal relation given a certain cue phrase (e.g., causative verb). Rink~et~al.~\cite{rink10} propose a Support Vector Machines classifier trained on contextual features. More recent approaches~\cite{dasgupta-etal-2018-automatic-extraction,li19} apply DL to automatically extract useful features from raw text. However, the existing approaches are not capable of extracting conditionals in fine-grained form, rendering them unsuitable for our use case. Specifically, some approaches extract only causal pairs or detect causal relations on the phrase level, but do not consider the combinatorics between causes and effects. In addition, they do not allow to split causes and effects into more granular text fragments (e.g., variable and condition). We addressed this research gap in previous work~\cite{fischbachAIRE21} and trained a Recursive Neural Tensor Network (RNTN) to recover the composition of conditionals as binary trees. However, we found that the RNTN struggles to understand the semantics of out-of-vocabulary words posing a threat to its applicability in practice. Contrary, \textsf{CiRA} is able to handle unseen data due to subword tokenization.

\section{Conclusion}
\label{sec:conclusion}
Acceptance testing evaluates the conformance between actual and expected system behavior. However, the creation of acceptance tests is laborious and requires manual work due to missing tool support. In this paper, we focus on conditional statements in functional requirements and demonstrate how NLP can be used to automatically generate the minimum set of required test cases. Specifically, we present our tool-supported approach \textsf{CiRA} capable of (1) detecting conditional statements, (2) extracting them in fine-grained form, and (3) translating them into a CEG, from which we derive the minimal number of required test cases. We evaluate \textsf{CiRA} by conducting a case study with three companies. Our study demonstrates that \textsf{CiRA} is able to automatically create 71.8~\% of the 578 manually created test cases. Additionally, \textsf{CiRA} identified 80 relevant test cases that were missed in manual test design. Future research will both explore the integration of \textsf{CiRA} in an industrial context, but also explore how the \textsf{CiRA} approach can support further use cases. By providing alternative representations of causal requirements -- both in the form of a CEG and a test suite -- potential improvements to the understandability of requirements can be investigated. For example, \textsf{CiRA} could be used to create an immediate feedback loop to confirm that these alternative representations reflect the original intention of the possibly ambiguous natural language requirement.

\paragraph{Data Availability} A live demo of \textsf{CiRA} can be accessed at \url{www.cira.bth.se/demo/}. Our code, annotated data sets, and all trained models are publicly available at \doi{10.5281/zenodo.5550387}.

\printcredits

\section*{Declaration of competing interest}
The authors declare that they have no known competing financial interests or personal relationships that could have appeared to influence the work reported in this paper.

\section*{Acknowledgments}
We would like to acknowledge that this work was supported by the KKS foundation through the S.E.R.T. Research Profile project at Blekinge Institute of Technology. We thank all collaborating companies for providing access to their data and test designers who volunteered their time to participate in this research.

\bibliographystyle{cas-model2-names}

\bibliography{cas-refs}

\begin{thebibliography}{57}
\expandafter\ifx\csname natexlab\endcsname\relax\def\natexlab#1{#1}\fi
\providecommand{\url}[1]{\texttt{#1}}
\providecommand{\href}[2]{#2}
\providecommand{\path}[1]{#1}
\providecommand{\DOIprefix}{doi:}
\providecommand{\ArXivprefix}{arXiv:}
\providecommand{\URLprefix}{URL: }
\providecommand{\Pubmedprefix}{pmid:}
\providecommand{\doi}[1]{\href{http://dx.doi.org/#1}{\path{#1}}}
\providecommand{\Pubmed}[1]{\href{pmid:#1}{\path{#1}}}
\providecommand{\bibinfo}[2]{#2}
\ifx\xfnm\relax \def\xfnm[#1]{\unskip,\space#1}\fi
\bibitem[{Ahsan et~al.(2017)Ahsan, Butt, Ahmed and Anwar}]{Ahsan17}
\bibinfo{author}{Ahsan, I.}, \bibinfo{author}{Butt, W.H.},
  \bibinfo{author}{Ahmed, M.A.}, \bibinfo{author}{Anwar, M.W.},
  \bibinfo{year}{2017}.
\newblock \bibinfo{title}{A comprehensive investigation of natural language
  processing techniques and tools to generate automated test cases}, in:
  \bibinfo{booktitle}{Proceedings of the Second International Conference on
  Internet of Things, Data and Cloud Computing},
  \bibinfo{publisher}{Association for Computing Machinery},
  \bibinfo{address}{New York, NY, USA}. p. \bibinfo{pages}{1–10}.
\newblock \URLprefix \url{https://doi.org/10.1145/3018896.3036375},
  \DOIprefix\doi{10.1145/3018896.3036375}.
\bibitem[{Barros et~al.(2011)Barros, Neves, Hori and Torres}]{barros11}
\bibinfo{author}{Barros, F.A.}, \bibinfo{author}{Neves, L.},
  \bibinfo{author}{Hori, E.}, \bibinfo{author}{Torres, D.},
  \bibinfo{year}{2011}.
\newblock \bibinfo{title}{{The ucsCNL: A Controlled Natural Language for Use
  Case Specifications}}, in: \bibinfo{booktitle}{{Proceedings of the 23rd
  International Conference on Software Engineering \& Knowledge Engineering
  (SEKE)}}, pp. \bibinfo{pages}{250--253}.
\bibitem[{Beller et~al.(2015)Beller, Gousios, Panichella and
  Zaidman}]{beller15}
\bibinfo{author}{Beller, M.}, \bibinfo{author}{Gousios, G.},
  \bibinfo{author}{Panichella, A.}, \bibinfo{author}{Zaidman, A.},
  \bibinfo{year}{2015}.
\newblock \bibinfo{title}{When, how, and why developers (do not) test in their
  ides}, in: \bibinfo{booktitle}{Proceedings of the 2015 10th Joint Meeting on
  Foundations of Software Engineering}, \bibinfo{publisher}{Association for
  Computing Machinery}, \bibinfo{address}{New York, NY, USA}. p.
  \bibinfo{pages}{179–190}.
\newblock \URLprefix \url{https://doi.org/10.1145/2786805.2786843},
  \DOIprefix\doi{10.1145/2786805.2786843}.
\bibitem[{Bergstra et~al.(2011)Bergstra, Bardenet, Bengio and
  K\'{e}gl}]{bergstra11}
\bibinfo{author}{Bergstra, J.}, \bibinfo{author}{Bardenet, R.},
  \bibinfo{author}{Bengio, Y.}, \bibinfo{author}{K\'{e}gl, B.},
  \bibinfo{year}{2011}.
\newblock \bibinfo{title}{Algorithms for hyper-parameter optimization}, in:
  \bibinfo{booktitle}{Proceedings of the 24th International Conference on
  Neural Information Processing Systems}, \bibinfo{publisher}{Curran Associates
  Inc.}, \bibinfo{address}{Red Hook, NY, USA}. p. \bibinfo{pages}{2546–2554}.
\bibitem[{Carvalho et~al.(2014)Carvalho, Barros, Lapschies, Schulze and
  Peleska}]{Carvalho14}
\bibinfo{author}{Carvalho, G.}, \bibinfo{author}{Barros, F.},
  \bibinfo{author}{Lapschies, F.}, \bibinfo{author}{Schulze, U.},
  \bibinfo{author}{Peleska, J.}, \bibinfo{year}{2014}.
\newblock \bibinfo{title}{Model-based testing from controlled natural language
  requirements}, in: \bibinfo{editor}{Artho, C.},
  \bibinfo{editor}{{\"O}lveczky, P.C.} (Eds.), \bibinfo{booktitle}{Formal
  Techniques for Safety-Critical Systems}, \bibinfo{publisher}{Springer
  International Publishing}, \bibinfo{address}{Cham}. pp.
  \bibinfo{pages}{19--35}.
\bibitem[{Chang and Choi(2005)}]{chang05}
\bibinfo{author}{Chang, D.S.}, \bibinfo{author}{Choi, K.S.},
  \bibinfo{year}{2005}.
\newblock \bibinfo{title}{Causal relation extraction using cue phrase and
  lexical pair probabilities}, in: \bibinfo{editor}{Su, K.Y.},
  \bibinfo{editor}{Tsujii, J.}, \bibinfo{editor}{Lee, J.H.},
  \bibinfo{editor}{Kwong, O.Y.} (Eds.), \bibinfo{booktitle}{Natural Language
  Processing -- IJCNLP 2004}, \bibinfo{publisher}{Springer Berlin Heidelberg},
  \bibinfo{address}{Berlin, Heidelberg}. pp. \bibinfo{pages}{61--70}.
\bibitem[{Dasgupta et~al.(2018)Dasgupta, Saha, Dey and
  Naskar}]{dasgupta-etal-2018-automatic-extraction}
\bibinfo{author}{Dasgupta, T.}, \bibinfo{author}{Saha, R.},
  \bibinfo{author}{Dey, L.}, \bibinfo{author}{Naskar, A.},
  \bibinfo{year}{2018}.
\newblock \bibinfo{title}{Automatic extraction of causal relations from text
  using linguistically informed deep neural networks}, in:
  \bibinfo{booktitle}{Proceedings of the 19th Annual {SIG}dial Meeting on
  Discourse and Dialogue}, \bibinfo{publisher}{Association for Computational
  Linguistics}, \bibinfo{address}{Melbourne, Australia}. pp.
  \bibinfo{pages}{306--316}.
\newblock \URLprefix \url{https://aclanthology.org/W18-5035},
  \DOIprefix\doi{10.18653/v1/W18-5035}.
\bibitem[{Devlin et~al.(2019)Devlin, Chang, Lee and Toutanova}]{bert}
\bibinfo{author}{Devlin, J.}, \bibinfo{author}{Chang, M.W.},
  \bibinfo{author}{Lee, K.}, \bibinfo{author}{Toutanova, K.},
  \bibinfo{year}{2019}.
\newblock \bibinfo{title}{{BERT}: Pre-training of deep bidirectional
  transformers for language understanding}, in: \bibinfo{booktitle}{Proceedings
  of the 2019 Conference of the North {A}merican Chapter of the Association for
  Computational Linguistics: Human Language Technologies, Volume 1 (Long and
  Short Papers)}, \bibinfo{publisher}{Association for Computational
  Linguistics}, \bibinfo{address}{Minneapolis, Minnesota}. pp.
  \bibinfo{pages}{4171--4186}.
\newblock \URLprefix \url{https://aclanthology.org/N19-1423},
  \DOIprefix\doi{10.18653/v1/N19-1423}.
\bibitem[{Dwarakanath and Sengupta(2012)}]{Dwarakanath12}
\bibinfo{author}{Dwarakanath, A.}, \bibinfo{author}{Sengupta, S.},
  \bibinfo{year}{2012}.
\newblock \bibinfo{title}{Litmus: Generation of test cases from functional
  requirements in natural language}, in: \bibinfo{editor}{Bouma, G.},
  \bibinfo{editor}{Ittoo, A.}, \bibinfo{editor}{M{\'e}tais, E.},
  \bibinfo{editor}{Wortmann, H.} (Eds.), \bibinfo{booktitle}{Natural Language
  Processing and Information Systems}, \bibinfo{publisher}{Springer Berlin
  Heidelberg}, \bibinfo{address}{Berlin, Heidelberg}. pp.
  \bibinfo{pages}{58--69}.
\bibitem[{Femmer et~al.(2017)Femmer, {Méndez Fernández}, Wagner and
  Eder}]{FEMMER2017190}
\bibinfo{author}{Femmer, H.}, \bibinfo{author}{{Méndez Fernández}, D.},
  \bibinfo{author}{Wagner, S.}, \bibinfo{author}{Eder, S.},
  \bibinfo{year}{2017}.
\newblock \bibinfo{title}{Rapid quality assurance with requirements smells}.
\newblock \bibinfo{journal}{Journal of Systems and Software}
  \bibinfo{volume}{123}, \bibinfo{pages}{190--213}.
\newblock \URLprefix
  \url{https://www.sciencedirect.com/science/article/pii/S0164121216000789},
  \DOIprefix\doi{https://doi.org/10.1016/j.jss.2016.02.047}.
\bibitem[{Fern\'{a}ndez et~al.(2017)Fern\'{a}ndez, Wagner, Kalinowski,
  Felderer, Mafra, Vetr\`{o}, Conte, Christiansson, Greer, Lassenius,
  M\"{a}nnist\"{o}, Nayabi, Oivo, Penzenstadler, Pfahl, Prikladnicki, Ruhe,
  Schekelmann, Sen, Spinola, Tuzcu, De~La~Vara and Wieringa}]{mendez}
\bibinfo{author}{Fern\'{a}ndez, D.M.}, \bibinfo{author}{Wagner, S.},
  \bibinfo{author}{Kalinowski, M.}, \bibinfo{author}{Felderer, M.},
  \bibinfo{author}{Mafra, P.}, \bibinfo{author}{Vetr\`{o}, A.},
  \bibinfo{author}{Conte, T.}, \bibinfo{author}{Christiansson, M.T.},
  \bibinfo{author}{Greer, D.}, \bibinfo{author}{Lassenius, C.},
  \bibinfo{author}{M\"{a}nnist\"{o}, T.}, \bibinfo{author}{Nayabi, M.},
  \bibinfo{author}{Oivo, M.}, \bibinfo{author}{Penzenstadler, B.},
  \bibinfo{author}{Pfahl, D.}, \bibinfo{author}{Prikladnicki, R.},
  \bibinfo{author}{Ruhe, G.}, \bibinfo{author}{Schekelmann, A.},
  \bibinfo{author}{Sen, S.}, \bibinfo{author}{Spinola, R.},
  \bibinfo{author}{Tuzcu, A.}, \bibinfo{author}{De~La~Vara, J.L.},
  \bibinfo{author}{Wieringa, R.}, \bibinfo{year}{2017}.
\newblock \bibinfo{title}{Naming the pain in requirements engineering}.
\newblock \bibinfo{journal}{Empirical Softw. Engg.} \bibinfo{volume}{22},
  \bibinfo{pages}{2298–2338}.
\bibitem[{Ferrari et~al.(2017)Ferrari, Spagnolo and Gnesi}]{Ferrari17}
\bibinfo{author}{Ferrari, A.}, \bibinfo{author}{Spagnolo, G.O.},
  \bibinfo{author}{Gnesi, S.}, \bibinfo{year}{2017}.
\newblock \bibinfo{title}{Pure: A dataset of public requirements documents},
  in: \bibinfo{booktitle}{2017 IEEE 25th International Requirements Engineering
  Conference (RE)}, pp. \bibinfo{pages}{502--505}.
\newblock \DOIprefix\doi{10.1109/RE.2017.29}.
\bibitem[{Fischbach et~al.(2020a)Fischbach, Femmer, Mendez, Fucci and
  Vogelsang}]{fischbachESEM}
\bibinfo{author}{Fischbach, J.}, \bibinfo{author}{Femmer, H.},
  \bibinfo{author}{Mendez, D.}, \bibinfo{author}{Fucci, D.},
  \bibinfo{author}{Vogelsang, A.}, \bibinfo{year}{2020}a.
\newblock \bibinfo{title}{What makes agile test artifacts useful? an
  activity-based quality model from a practitioners' perspective}, in:
  \bibinfo{booktitle}{Proceedings of the 14th ACM / IEEE International
  Symposium on Empirical Software Engineering and Measurement (ESEM)},
  \bibinfo{publisher}{Association for Computing Machinery},
  \bibinfo{address}{New York, NY, USA}. pp. \bibinfo{pages}{1--10}.
\newblock \URLprefix \url{https://doi.org/10.1145/3382494.3421462},
  \DOIprefix\doi{10.1145/3382494.3421462}.
\bibitem[{Fischbach et~al.(2021a)Fischbach, Frattini, Mendez, Unterkalmsteiner,
  Femmer and Vogelsang}]{fischbach2021profes}
\bibinfo{author}{Fischbach, J.}, \bibinfo{author}{Frattini, J.},
  \bibinfo{author}{Mendez, D.}, \bibinfo{author}{Unterkalmsteiner, M.},
  \bibinfo{author}{Femmer, H.}, \bibinfo{author}{Vogelsang, A.},
  \bibinfo{year}{2021}a.
\newblock \bibinfo{title}{How do practitioners interpret conditionals in
  requirements?}, in: \bibinfo{editor}{Ardito, L.},
  \bibinfo{editor}{Jedlitschka, A.}, \bibinfo{editor}{Morisio, M.},
  \bibinfo{editor}{Torchiano, M.} (Eds.), \bibinfo{booktitle}{Product-Focused
  Software Process Improvement}, \bibinfo{publisher}{Springer International
  Publishing}, \bibinfo{address}{Cham}. pp. \bibinfo{pages}{85--102}.
\bibitem[{Fischbach et~al.(2021b)Fischbach, Frattini, Spaans, Kummeth,
  Vogelsang, Mendez and Unterkalmsteiner}]{fischbachREFSQ}
\bibinfo{author}{Fischbach, J.}, \bibinfo{author}{Frattini, J.},
  \bibinfo{author}{Spaans, A.}, \bibinfo{author}{Kummeth, M.},
  \bibinfo{author}{Vogelsang, A.}, \bibinfo{author}{Mendez, D.},
  \bibinfo{author}{Unterkalmsteiner, M.}, \bibinfo{year}{2021}b.
\newblock \bibinfo{title}{Automatic detection of causality in requirement
  artifacts: The cira approach}, in: \bibinfo{editor}{Dalpiaz, F.},
  \bibinfo{editor}{Spoletini, P.} (Eds.), \bibinfo{booktitle}{Requirements
  Engineering: Foundation for Software Quality}, \bibinfo{publisher}{Springer
  International Publishing}, \bibinfo{address}{Cham}. pp.
  \bibinfo{pages}{19--36}.
\bibitem[{Fischbach et~al.(2020b)Fischbach, Hauptmann, Konwitschny, Spies and
  Vogelsang}]{fischbachRENEXT}
\bibinfo{author}{Fischbach, J.}, \bibinfo{author}{Hauptmann, B.},
  \bibinfo{author}{Konwitschny, L.}, \bibinfo{author}{Spies, D.},
  \bibinfo{author}{Vogelsang, A.}, \bibinfo{year}{2020}b.
\newblock \bibinfo{title}{Towards causality extraction from requirements}, in:
  \bibinfo{booktitle}{2020 IEEE 28th International Requirements Engineering
  Conference (RE)}, pp. \bibinfo{pages}{388--393}.
\newblock \DOIprefix\doi{10.1109/RE48521.2020.00053}.
\bibitem[{Fischbach et~al.(2021c)Fischbach, Springer, Frattini, Femmer,
  Vogelsang and Mendez}]{fischbachAIRE21}
\bibinfo{author}{Fischbach, J.}, \bibinfo{author}{Springer, T.},
  \bibinfo{author}{Frattini, J.}, \bibinfo{author}{Femmer, H.},
  \bibinfo{author}{Vogelsang, A.}, \bibinfo{author}{Mendez, D.},
  \bibinfo{year}{2021}c.
\newblock \bibinfo{title}{Fine-grained causality extraction from natural
  language requirements using recursive neural tensor networks}, in:
  \bibinfo{booktitle}{2021 IEEE 29th International Requirements Engineering
  Conference Workshops (REW)}, pp. \bibinfo{pages}{60--69}.
\newblock \DOIprefix\doi{10.1109/REW53955.2021.00016}.
\bibitem[{Fischbach et~al.(2020c)Fischbach, Vogelsang, Spies, Wehrle, Junker
  and Freudenstein}]{fischbachICST}
\bibinfo{author}{Fischbach, J.}, \bibinfo{author}{Vogelsang, A.},
  \bibinfo{author}{Spies, D.}, \bibinfo{author}{Wehrle, A.},
  \bibinfo{author}{Junker, M.}, \bibinfo{author}{Freudenstein, D.},
  \bibinfo{year}{2020}c.
\newblock \bibinfo{title}{Specmate: Automated creation of test cases from
  acceptance criteria}, in: \bibinfo{booktitle}{2020 IEEE 13th International
  Conference on Software Testing, Validation and Verification (ICST)}, pp.
  \bibinfo{pages}{321--331}.
\newblock \DOIprefix\doi{10.1109/ICST46399.2020.00040}.
\bibitem[{Garousi et~al.(2020)Garousi, Bauer and Felderer}]{GAROUSI}
\bibinfo{author}{Garousi, V.}, \bibinfo{author}{Bauer, S.},
  \bibinfo{author}{Felderer, M.}, \bibinfo{year}{2020}.
\newblock \bibinfo{title}{Nlp-assisted software testing: A systematic mapping
  of the literature}.
\newblock \bibinfo{journal}{Information and Software Technology}
  \bibinfo{volume}{126}, \bibinfo{pages}{106321}.
\newblock \URLprefix
  \url{https://www.sciencedirect.com/science/article/pii/S0950584920300744},
  \DOIprefix\doi{https://doi.org/10.1016/j.infsof.2020.106321}.
\bibitem[{Girju et~al.(2007)Girju, Nakov, Nastase, Szpakowicz, Turney and
  Yuret}]{semeval07}
\bibinfo{author}{Girju, R.}, \bibinfo{author}{Nakov, P.},
  \bibinfo{author}{Nastase, V.}, \bibinfo{author}{Szpakowicz, S.},
  \bibinfo{author}{Turney, P.}, \bibinfo{author}{Yuret, D.},
  \bibinfo{year}{2007}.
\newblock \bibinfo{title}{{S}em{E}val-2007 task 04: Classification of semantic
  relations between nominals}, in: \bibinfo{booktitle}{Proceedings of the
  Fourth International Workshop on Semantic Evaluations ({S}em{E}val-2007)},
  \bibinfo{publisher}{Association for Computational Linguistics},
  \bibinfo{address}{Prague, Czech Republic}. pp. \bibinfo{pages}{13--18}.
\newblock \URLprefix \url{https://aclanthology.org/S07-1003}.
\bibitem[{Goffi et~al.(2016)Goffi, Gorla, Ernst and Pezz{\`{e}}}]{Goffi2016}
\bibinfo{author}{Goffi, A.}, \bibinfo{author}{Gorla, A.},
  \bibinfo{author}{Ernst, M.D.}, \bibinfo{author}{Pezz{\`{e}}, M.},
  \bibinfo{year}{2016}.
\newblock \bibinfo{title}{Automatic generation of oracles for exceptional
  behaviors}, in: \bibinfo{booktitle}{Proceedings of the 25th International
  Symposium on Software Testing and Analysis}, \bibinfo{publisher}{{ACM}}.
\newblock \URLprefix \url{https://doi.org/10.1145/2931037.2931061},
  \DOIprefix\doi{10.1145/2931037.2931061}.
\bibitem[{Gülle et~al.(2020)Gülle, Ford, Ebel, Brokhausen and
  Vogelsang}]{Guelle20}
\bibinfo{author}{Gülle, K.J.}, \bibinfo{author}{Ford, N.},
  \bibinfo{author}{Ebel, P.}, \bibinfo{author}{Brokhausen, F.},
  \bibinfo{author}{Vogelsang, A.}, \bibinfo{year}{2020}.
\newblock \bibinfo{title}{Topic modeling on user stories using word mover's
  distance}, in: \bibinfo{booktitle}{2020 IEEE Seventh International Workshop
  on Artificial Intelligence for Requirements Engineering (AIRE)}, pp.
  \bibinfo{pages}{52--60}.
\newblock \DOIprefix\doi{10.1109/AIRE51212.2020.00015}.
\bibitem[{Hendrickx et~al.(2010)Hendrickx, Kim, Kozareva, Nakov,
  {\'O}~S{\'e}aghdha, Pad{\'o}, Pennacchiotti, Romano and
  Szpakowicz}]{semeval10}
\bibinfo{author}{Hendrickx, I.}, \bibinfo{author}{Kim, S.N.},
  \bibinfo{author}{Kozareva, Z.}, \bibinfo{author}{Nakov, P.},
  \bibinfo{author}{{\'O}~S{\'e}aghdha, D.}, \bibinfo{author}{Pad{\'o}, S.},
  \bibinfo{author}{Pennacchiotti, M.}, \bibinfo{author}{Romano, L.},
  \bibinfo{author}{Szpakowicz, S.}, \bibinfo{year}{2010}.
\newblock \bibinfo{title}{{S}em{E}val-2010 task 8: Multi-way classification of
  semantic relations between pairs of nominals}, in:
  \bibinfo{booktitle}{Proceedings of the 5th International Workshop on Semantic
  Evaluation}, \bibinfo{publisher}{Association for Computational Linguistics},
  \bibinfo{address}{Uppsala, Sweden}. pp. \bibinfo{pages}{33--38}.
\newblock \URLprefix \url{https://aclanthology.org/S10-1006}.
\bibitem[{Hey et~al.(2020)Hey, Keim, Koziolek and Tichy}]{Hey20}
\bibinfo{author}{Hey, T.}, \bibinfo{author}{Keim, J.},
  \bibinfo{author}{Koziolek, A.}, \bibinfo{author}{Tichy, W.F.},
  \bibinfo{year}{2020}.
\newblock \bibinfo{title}{Norbert: Transfer learning for requirements
  classification}, in: \bibinfo{booktitle}{2020 IEEE 28th International
  Requirements Engineering Conference (RE)}, pp. \bibinfo{pages}{169--179}.
\newblock \DOIprefix\doi{10.1109/RE48521.2020.00028}.
\bibitem[{Hripcsak(2005)}]{hripcsak05}
\bibinfo{author}{Hripcsak, G.}, \bibinfo{year}{2005}.
\newblock \bibinfo{title}{Agreement, the f-measure, and reliability in
  information retrieval}.
\newblock \bibinfo{journal}{Journal of the American Medical Informatics
  Association} \bibinfo{volume}{12}, \bibinfo{pages}{296--298}.
\newblock \URLprefix \url{https://doi.org/10.1197/jamia.m1733},
  \DOIprefix\doi{10.1197/jamia.m1733}.
\bibitem[{ISO/IEC/IEEE 24765:2010(E)()}]{ISONorm}
ISO/IEC/IEEE 24765:2010(E), \bibinfo{year}{2011}.
\newblock \bibinfo{title}{{Systems and software engineering — Vocabulary}}.
\newblock \bibinfo{type}{Standard}. International Organization for
  Standardization. \bibinfo{address}{Geneva, CH}.
\bibitem[{James et~al.(2013)James, Witten, Hastie and Tibshirani}]{James13}
\bibinfo{author}{James, G.}, \bibinfo{author}{Witten, D.},
  \bibinfo{author}{Hastie, T.}, \bibinfo{author}{Tibshirani, R.E.},
  \bibinfo{year}{2013}.
\newblock \bibinfo{title}{An Introduction to Statistical Learning}. volume
  \bibinfo{volume}{112}.
\newblock \bibinfo{publisher}{Springer}.
\bibitem[{Kassab et~al.(2014)Kassab, Neill and Laplante}]{Kassab14}
\bibinfo{author}{Kassab, M.}, \bibinfo{author}{Neill, C.},
  \bibinfo{author}{Laplante, P.}, \bibinfo{year}{2014}.
\newblock \bibinfo{title}{State of practice in requirements engineering:
  contemporary data}.
\newblock \bibinfo{journal}{Innovations in Systems and Software Engineering}
  \bibinfo{volume}{10}, \bibinfo{pages}{235--241}.
\newblock \URLprefix \url{https://doi.org/10.1007/s11334-014-0232-4},
  \DOIprefix\doi{10.1007/s11334-014-0232-4}.
\bibitem[{Khoo et~al.(1998)Khoo, Kornfilt, Oddy and Myaeng}]{khoo98}
\bibinfo{author}{Khoo, C.S.G.}, \bibinfo{author}{Kornfilt, J.},
  \bibinfo{author}{Oddy, R.N.}, \bibinfo{author}{Myaeng, S.H.},
  \bibinfo{year}{1998}.
\newblock \bibinfo{title}{Automatic extraction of cause-effect information from
  newspaper text without knowledge-based inferencing}.
\newblock \bibinfo{journal}{Literary and Linguistic Computing}
  \bibinfo{volume}{13}, \bibinfo{pages}{177--186}.
\newblock \URLprefix \url{https://doi.org/10.1093/llc/13.4.177},
  \DOIprefix\doi{10.1093/llc/13.4.177}.
\bibitem[{Kitchenham and Pfleeger(2002)}]{Kitchenham}
\bibinfo{author}{Kitchenham, B.}, \bibinfo{author}{Pfleeger, S.L.},
  \bibinfo{year}{2002}.
\newblock \bibinfo{title}{Principles of survey research: Part 5: Populations
  and samples}.
\newblock \bibinfo{journal}{SIGSOFT Softw. Eng. Notes} \bibinfo{volume}{27},
  \bibinfo{pages}{17–20}.
\newblock \URLprefix \url{https://doi.org/10.1145/571681.571686},
  \DOIprefix\doi{10.1145/571681.571686}.
\bibitem[{Kolditz et~al.(2019)Kolditz, Lohr, Hellrich, Modersohn, Betz,
  Kiehntopf and Hahn}]{kolditz19}
\bibinfo{author}{Kolditz, T.}, \bibinfo{author}{Lohr, C.},
  \bibinfo{author}{Hellrich, J.}, \bibinfo{author}{Modersohn, L.},
  \bibinfo{author}{Betz, B.}, \bibinfo{author}{Kiehntopf, M.},
  \bibinfo{author}{Hahn, U.}, \bibinfo{year}{2019}.
\newblock \bibinfo{title}{Annotating german clinical documents for
  de-identification}.
\newblock \bibinfo{journal}{Studies in health technology and informatics}
  \bibinfo{volume}{264}, \bibinfo{pages}{203—207}.
\newblock \URLprefix \url{https://doi.org/10.3233/SHTI190212},
  \DOIprefix\doi{10.3233/shti190212}.
\bibitem[{Kumar et~al.(2020)Kumar, Makhija and Gupta}]{kumar-etal-2020-noisy}
\bibinfo{author}{Kumar, A.}, \bibinfo{author}{Makhija, P.},
  \bibinfo{author}{Gupta, A.}, \bibinfo{year}{2020}.
\newblock \bibinfo{title}{Noisy text data: Achilles{'} heel of {BERT}}, in:
  \bibinfo{booktitle}{Proceedings of the Sixth Workshop on Noisy User-generated
  Text (W-NUT 2020)}, \bibinfo{publisher}{Association for Computational
  Linguistics}, \bibinfo{address}{Online}. pp. \bibinfo{pages}{16--21}.
\newblock \URLprefix \url{https://aclanthology.org/2020.wnut-1.3},
  \DOIprefix\doi{10.18653/v1/2020.wnut-1.3}.
\bibitem[{Li et~al.(2021)Li, Li, Zou and Ren}]{li19}
\bibinfo{author}{Li, Z.}, \bibinfo{author}{Li, Q.}, \bibinfo{author}{Zou, X.},
  \bibinfo{author}{Ren, J.}, \bibinfo{year}{2021}.
\newblock \bibinfo{title}{Causality extraction based on self-attentive
  bilstm-crf with transferred embeddings}.
\newblock \bibinfo{journal}{Neurocomputing} \bibinfo{volume}{423},
  \bibinfo{pages}{207--219}.
\newblock \DOIprefix\doi{https://doi.org/10.1016/j.neucom.2020.08.078}.
\bibitem[{Liu and Nakajima(2020)}]{shaoying}
\bibinfo{author}{Liu, S.}, \bibinfo{author}{Nakajima, S.},
  \bibinfo{year}{2020}.
\newblock \bibinfo{title}{Automatic test case and test oracle generation based
  on functional scenarios in formal specifications for conformance testing}.
\newblock \bibinfo{journal}{IEEE Transactions on Software Engineering} ,
  \bibinfo{pages}{1--1}\DOIprefix\doi{10.1109/TSE.2020.2999884}.
\bibitem[{Liu et~al.(2019)Liu, Ott, Goyal, Du, Joshi, Chen, Levy, Lewis,
  Zettlemoyer and Stoyanov}]{roberta}
\bibinfo{author}{Liu, Y.}, \bibinfo{author}{Ott, M.}, \bibinfo{author}{Goyal,
  N.}, \bibinfo{author}{Du, J.}, \bibinfo{author}{Joshi, M.},
  \bibinfo{author}{Chen, D.}, \bibinfo{author}{Levy, O.},
  \bibinfo{author}{Lewis, M.}, \bibinfo{author}{Zettlemoyer, L.},
  \bibinfo{author}{Stoyanov, V.}, \bibinfo{year}{2019}.
\newblock \bibinfo{title}{Roberta: {A} robustly optimized {BERT} pretraining
  approach}.
\newblock \bibinfo{journal}{CoRR} \bibinfo{volume}{abs/1907.11692}.
\bibitem[{Mikolov et~al.(2013)Mikolov, Sutskever, Chen, Corrado and
  Dean}]{mikolov13}
\bibinfo{author}{Mikolov, T.}, \bibinfo{author}{Sutskever, I.},
  \bibinfo{author}{Chen, K.}, \bibinfo{author}{Corrado, G.},
  \bibinfo{author}{Dean, J.}, \bibinfo{year}{2013}.
\newblock \bibinfo{title}{Distributed representations of words and phrases and
  their compositionality}, in: \bibinfo{booktitle}{Proceedings of the 26th
  International Conference on Neural Information Processing Systems - Volume
  2}, \bibinfo{publisher}{Curran Associates Inc.}, \bibinfo{address}{Red Hook,
  NY, USA}. p. \bibinfo{pages}{3111–3119}.
\bibitem[{Myers et~al.(2012)Myers, Badgett and Sandler}]{Myers}
\bibinfo{editor}{Myers, G.J.}, \bibinfo{editor}{Badgett, T.},
  \bibinfo{editor}{Sandler, C.} (Eds.), \bibinfo{year}{2012}.
\newblock \bibinfo{title}{The Art of Software Testing}.
\newblock \bibinfo{publisher}{Wiley}.
\newblock \URLprefix \url{https://doi.org/10.1002/9781119202486},
  \DOIprefix\doi{10.1002/9781119202486}.
\bibitem[{Nursimulu and Probert(1995)}]{Nursimulu95}
\bibinfo{author}{Nursimulu, K.}, \bibinfo{author}{Probert, R.L.},
  \bibinfo{year}{1995}.
\newblock \bibinfo{title}{Cause-effect graphing analysis and validation of
  requirements}, in: \bibinfo{booktitle}{Proceedings of the 1995 Conference of
  the Centre for Advanced Studies on Collaborative Research},
  \bibinfo{publisher}{IBM Press}. p.~\bibinfo{pages}{46}.
\bibitem[{Puente and Olivas(2008)}]{puente}
\bibinfo{author}{Puente, C.}, \bibinfo{author}{Olivas, J.},
  \bibinfo{year}{2008}.
\newblock \bibinfo{title}{Analysis, detection and classification of certain
  conditional sentences in text documents}, in:
  \bibinfo{booktitle}{International Conference on Information Processing and
  Management of Uncertainty in Knowledge-Based Systems}, p.
  \bibinfo{pages}{1097–1104}.
\bibitem[{Rink and Harabagiu(2010)}]{rink10}
\bibinfo{author}{Rink, B.}, \bibinfo{author}{Harabagiu, S.},
  \bibinfo{year}{2010}.
\newblock \bibinfo{title}{{UTD}: Classifying semantic relations by combining
  lexical and semantic resources}, in: \bibinfo{booktitle}{Proceedings of the
  5th International Workshop on Semantic Evaluation},
  \bibinfo{publisher}{Association for Computational Linguistics},
  \bibinfo{address}{Uppsala, Sweden}. pp. \bibinfo{pages}{256--259}.
\newblock \URLprefix \url{https://aclanthology.org/S10-1057}.
\bibitem[{Runeson and H\"{o}st(2009)}]{runeson}
\bibinfo{author}{Runeson, P.}, \bibinfo{author}{H\"{o}st, M.},
  \bibinfo{year}{2009}.
\newblock \bibinfo{title}{Guidelines for conducting and reporting case study
  research in software engineering}.
\newblock \bibinfo{journal}{Empirical Softw. Engg.} \bibinfo{volume}{14},
  \bibinfo{pages}{131–164}.
\bibitem[{Sanh et~al.(2019)Sanh, Debut, Chaumond and Wolf}]{distillbert}
\bibinfo{author}{Sanh, V.}, \bibinfo{author}{Debut, L.},
  \bibinfo{author}{Chaumond, J.}, \bibinfo{author}{Wolf, T.},
  \bibinfo{year}{2019}.
\newblock \bibinfo{title}{Distilbert, a distilled version of {BERT:} smaller,
  faster, cheaper and lighter}.
\newblock \bibinfo{journal}{CoRR} \bibinfo{volume}{abs/1910.01108}.
\bibitem[{Santiago~J\'{u}nior and Vijaykumar(2012)}]{santiagoSpace}
\bibinfo{author}{Santiago~J\'{u}nior, V.A.d.}, \bibinfo{author}{Vijaykumar,
  V.L.}, \bibinfo{year}{2012}.
\newblock \bibinfo{title}{Generating model-based test cases from natural
  language requirements for space application software}.
\newblock \bibinfo{journal}{Software Quality Journal} \bibinfo{volume}{20},
  \bibinfo{pages}{77–143}.
\bibitem[{Sarmiento et~al.(2016)Sarmiento, Leite, Almentero and {Sotomayor
  Alzamora}}]{Sarmiento}
\bibinfo{author}{Sarmiento, E.}, \bibinfo{author}{Leite, J.C.},
  \bibinfo{author}{Almentero, E.}, \bibinfo{author}{{Sotomayor Alzamora}, G.},
  \bibinfo{year}{2016}.
\newblock \bibinfo{title}{Test scenario generation from natural language
  requirements descriptions based on petri-nets}.
\newblock \bibinfo{journal}{Electronic Notes in Theoretical Computer Science}
  \bibinfo{volume}{329}, \bibinfo{pages}{123--148}.
\bibitem[{Sharma and Biswas(2014)}]{enase14}
\bibinfo{author}{Sharma, R.}, \bibinfo{author}{Biswas, K.K.},
  \bibinfo{year}{2014}.
\newblock \bibinfo{title}{Automated generation of test cases from logical
  specification of software requirements}, in: \bibinfo{editor}{Filipe, J.},
  \bibinfo{editor}{Maciaszek, L.A.} (Eds.), \bibinfo{booktitle}{{ENASE} 2014 -
  Proceedings of the 9th International Conference on Evaluation of Novel
  Approaches to Software Engineering, Lisbon, Portugal, 28-30 April, 2014},
  \bibinfo{publisher}{SciTePress}. pp. \bibinfo{pages}{241--248}.
\newblock \URLprefix \url{https://doi.org/10.5220/0004972902410248},
  \DOIprefix\doi{10.5220/0004972902410248}.
\bibitem[{Sneed(2007)}]{Sneed07}
\bibinfo{author}{Sneed, H.M.}, \bibinfo{year}{2007}.
\newblock \bibinfo{title}{Testing against natural language requirements}, in:
  \bibinfo{booktitle}{2007 7th International Conference on Quality Software},
  \bibinfo{publisher}{IEEE Computer Society}, \bibinfo{address}{Los Alamitos,
  CA, USA}. pp. \bibinfo{pages}{380--387}.
\newblock \URLprefix
  \url{https://doi.ieeecomputersociety.org/10.1109/QSIC.2007.61},
  \DOIprefix\doi{10.1109/QSIC.2007.61}.
\bibitem[{Sorgente et~al.(2013)Sorgente, Vettigli and Mele}]{Sorgente2013}
\bibinfo{author}{Sorgente, A.}, \bibinfo{author}{Vettigli, G.},
  \bibinfo{author}{Mele, F.}, \bibinfo{year}{2013}.
\newblock \bibinfo{title}{Automatic extraction of cause-effect relations in
  natural language text}, in: \bibinfo{booktitle}{DART@AI*IA}, p.
  \bibinfo{pages}{37–48}.
\bibitem[{Stenetorp et~al.(2012)Stenetorp, Pyysalo, Topi{\'c}, Ohta, Ananiadou
  and Tsujii}]{brat12}
\bibinfo{author}{Stenetorp, P.}, \bibinfo{author}{Pyysalo, S.},
  \bibinfo{author}{Topi{\'c}, G.}, \bibinfo{author}{Ohta, T.},
  \bibinfo{author}{Ananiadou, S.}, \bibinfo{author}{Tsujii, J.},
  \bibinfo{year}{2012}.
\newblock \bibinfo{title}{brat: a web-based tool for {NLP}-assisted text
  annotation}, in: \bibinfo{booktitle}{Proceedings of the Demonstrations at the
  13th Conference of the {E}uropean Chapter of the Association for
  Computational Linguistics}, \bibinfo{publisher}{Association for Computational
  Linguistics}, \bibinfo{address}{Avignon, France}. pp.
  \bibinfo{pages}{102--107}.
\newblock \URLprefix \url{https://aclanthology.org/E12-2021}.
\bibitem[{Sundararaman et~al.(2019)Sundararaman, Subramanian, Wang, Si, Shen,
  Wang and Carin}]{sundararaman2019}
\bibinfo{author}{Sundararaman, D.}, \bibinfo{author}{Subramanian, V.},
  \bibinfo{author}{Wang, G.}, \bibinfo{author}{Si, S.}, \bibinfo{author}{Shen,
  D.}, \bibinfo{author}{Wang, D.}, \bibinfo{author}{Carin, L.},
  \bibinfo{year}{2019}.
\newblock \bibinfo{title}{Syntax-infused transformer and {BERT} models for
  machine translation and natural language understanding}.
\newblock \bibinfo{journal}{CoRR} \bibinfo{volume}{abs/1911.06156}.
\newblock \URLprefix \url{http://arxiv.org/abs/1911.06156},
  \href{http://arxiv.org/abs/1911.06156}{\tt arXiv:1911.06156}.
\bibitem[{Tran et~al.(2021)Tran, Unterkalmsteiner, Börstler and bin
  Ali}]{tran21}
\bibinfo{author}{Tran, H.K.V.}, \bibinfo{author}{Unterkalmsteiner, M.},
  \bibinfo{author}{Börstler, J.}, \bibinfo{author}{bin Ali, N.},
  \bibinfo{year}{2021}.
\newblock \bibinfo{title}{Assessing test artifact quality—a tertiary study}.
\newblock \bibinfo{journal}{Information and Software Technology}
  \bibinfo{volume}{139}, \bibinfo{pages}{106620}.
\newblock \URLprefix
  \url{https://www.sciencedirect.com/science/article/pii/S0950584921000938},
  \DOIprefix\doi{https://doi.org/10.1016/j.infsof.2021.106620}.
\bibitem[{Verma and Beg(2013)}]{Verma}
\bibinfo{author}{Verma, R.P.}, \bibinfo{author}{Beg, M.R.},
  \bibinfo{year}{2013}.
\newblock \bibinfo{title}{Generation of test cases from software requirements
  using natural language processing}, in: \bibinfo{booktitle}{2013 6th
  International Conference on Emerging Trends in Engineering and Technology},
  pp. \bibinfo{pages}{140--147}.
\newblock \DOIprefix\doi{10.1109/ICETET.2013.45}.
\bibitem[{Wang et~al.(2020)Wang, Pastore, Goknil and Briand}]{wangbriand}
\bibinfo{author}{Wang, C.}, \bibinfo{author}{Pastore, F.},
  \bibinfo{author}{Goknil, A.}, \bibinfo{author}{Briand, L.},
  \bibinfo{year}{2020}.
\newblock \bibinfo{title}{Automatic generation of acceptance test cases from
  use case specifications: an nlp-based approach}.
\newblock \bibinfo{journal}{IEEE Transactions on Software Engineering} ,
  \bibinfo{pages}{1--1}.
\bibitem[{Whalen et~al.(2006)Whalen, Rajan, Heimdahl and Miller}]{whalen06}
\bibinfo{author}{Whalen, M.W.}, \bibinfo{author}{Rajan, A.},
  \bibinfo{author}{Heimdahl, M.P.}, \bibinfo{author}{Miller, S.P.},
  \bibinfo{year}{2006}.
\newblock \bibinfo{title}{Coverage metrics for requirements-based testing}, in:
  \bibinfo{booktitle}{Proceedings of the 2006 International Symposium on
  Software Testing and Analysis}, \bibinfo{publisher}{Association for Computing
  Machinery}, \bibinfo{address}{New York, NY, USA}. p.
  \bibinfo{pages}{25–36}.
\newblock \URLprefix \url{https://doi.org/10.1145/1146238.1146242},
  \DOIprefix\doi{10.1145/1146238.1146242}.
\bibitem[{Xu et~al.(2020)Xu, Zuo, Liang and Zuo}]{xu-etal-2020-review}
\bibinfo{author}{Xu, J.}, \bibinfo{author}{Zuo, W.}, \bibinfo{author}{Liang,
  S.}, \bibinfo{author}{Zuo, X.}, \bibinfo{year}{2020}.
\newblock \bibinfo{title}{A review of dataset and labeling methods for
  causality extraction}, in: \bibinfo{booktitle}{Proceedings of the 28th
  International Conference on Computational Linguistics},
  \bibinfo{publisher}{International Committee on Computational Linguistics},
  \bibinfo{address}{Barcelona, Spain (Online)}. pp.
  \bibinfo{pages}{1519--1531}.
\newblock \URLprefix \url{https://aclanthology.org/2020.coling-main.133},
  \DOIprefix\doi{10.18653/v1/2020.coling-main.133}.
\bibitem[{Yang et~al.(2021)Yang, Han and Poon}]{yang2021survey}
\bibinfo{author}{Yang, J.}, \bibinfo{author}{Han, S.C.}, \bibinfo{author}{Poon,
  J.}, \bibinfo{year}{2021}.
\newblock \bibinfo{title}{A survey on extraction of causal relations from
  natural language text}.
\newblock \bibinfo{journal}{CoRR} \bibinfo{volume}{abs/2101.06426}.
\newblock \URLprefix \url{https://arxiv.org/abs/2101.06426},
  \href{http://arxiv.org/abs/2101.06426}{\tt arXiv:2101.06426}.
\bibitem[{Zhang et~al.(2014)Zhang, Yue, Ali, Zhang and Wu}]{Zhang14}
\bibinfo{author}{Zhang, M.}, \bibinfo{author}{Yue, T.}, \bibinfo{author}{Ali,
  S.}, \bibinfo{author}{Zhang, H.}, \bibinfo{author}{Wu, J.},
  \bibinfo{year}{2014}.
\newblock \bibinfo{title}{A systematic approach to automatically derive test
  cases from use cases specified in restricted natural languages}, in:
  \bibinfo{editor}{Amyot, D.}, \bibinfo{editor}{Fonseca~i Casas, P.},
  \bibinfo{editor}{Mussbacher, G.} (Eds.), \bibinfo{booktitle}{System Analysis
  and Modeling: Models and Reusability}, \bibinfo{publisher}{Springer
  International Publishing}, \bibinfo{address}{Cham}. pp.
  \bibinfo{pages}{142--157}.
\bibitem[{Zhao et~al.(2021)Zhao, Alhoshan, Ferrari, Letsholo, Ajagbe, Chioasca
  and Batista-Navarro}]{Zhao21}
\bibinfo{author}{Zhao, L.}, \bibinfo{author}{Alhoshan, W.},
  \bibinfo{author}{Ferrari, A.}, \bibinfo{author}{Letsholo, K.J.},
  \bibinfo{author}{Ajagbe, M.A.}, \bibinfo{author}{Chioasca, E.V.},
  \bibinfo{author}{Batista-Navarro, R.T.}, \bibinfo{year}{2021}.
\newblock \bibinfo{title}{Natural language processing for requirements
  engineering: A systematic mapping study}.
\newblock \bibinfo{journal}{ACM Comput. Surv.} \bibinfo{volume}{54}.
\newblock \URLprefix \url{https://doi.org/10.1145/3444689},
  \DOIprefix\doi{10.1145/3444689}.

\end{thebibliography}


\bio{}
\textbf{Jannik Fischbach} is a Ph.D. student at the Institute of Computer Science of the University of Cologne. From 2019 to 2022, Jannik also worked as a consultant at Qualicen GmbH - a spin-off founded out of the Technical University of Munich focusing on software and systems engineering. In June 2022, he joined Netlight as a consultant. His main research interests include requirements engineering and, in particular, the application of natural language processing methods on requirements artifacts. Jannik holds a Master’s degree in Information Systems from the Technical University of Munich.
\endbio

\bio{}
\textbf{Julian Frattini} is a Ph.D. student at the Blekinge Institute of Technology located in Karlskrona, Sweden. Since 2020 he is working under the supervision of Daniel Mendez in the area of requirements quality, specifically investigating the notion of good-enough requirements engineering. Through the collaboration with Ericsson Karlskrona the research is applied and grounded in practice. Julian holds a Master's degree in Informatics from the Technical University of Munich.
\endbio

\bio{}
\textbf{Andreas Vogelsang} is full professor for Software and Systems Engineering at the University of Cologne. He received a PhD from the Technical University of Munich. His research interests comprise requirements engineering, model-based systems engineering, and software architectures for embedded systems. He has published over 70 papers in international journals and conferences such as TSE, JSS, IEEE Software, and ICSE. In 2018, he was appointed as Junior-Fellow of the
German Society for Informatics (GI). Further information can be obtained from \url{https://cs.uni-koeln.de/sse}.
\endbio

\bio{}
\textbf{Daniel Mendez} is full professor at the Blekinge Institute of Technology, Sweden, and Lead Researcher heading the research division Requirements Engineering at fortiss, the research and transfer institute of the Free State of Bavaria for software-intensive systems and services. After studying Computer Science and Cognitive Neuroscience at the Ludwig Maximilian University of Munich, he pursued his doctoral and his habilitation degrees at the Technical University of Munich. His research is since then on Empirical Software Engineering with a particular focus on interdisciplinary, qualitative research in Requirements Engineering and its quality improvement – all in close collaboration with the relevant industries. He is further editorial board member for EMSE and JSS where he co-chairs the special tracks Reproducibility \& Open Science (EMSE) and In Practice (JSS) respectively. Finally, he is a member of the ACM, the German association of university professors and lecturers, the German Informatics Society, and ISERN. Further information is available at \url{http://www.mendezfe.org}.
\endbio

\bio{}
\textbf{Michael Unterkalmsteiner} is a senior lecturer at the Blekinge Institute of Technology, Sweden, where he also received a PhD in Software Engineering. He has been researching Software Engineering since 2009, focusing in particular on the coordination between requirements engineering and software testing. His research work is shaped by empirical problem identification, in-depth analysis of the state-of-art and practice, and collaborative solution development. This empirical, practice-driven approach has led to innovative and scalable solutions. His current research focuses on designing and implementing automated decision support systems for software engineers. Further information is available at \url{https://lmsteiner.com}.
\endbio

\bio{}
\textbf{Andreas Wehrle} is a software engineer at Allianz Deutschland AG. His main interest is the application of natural language processing methods to requirements. Andreas holds a Master’s degree in Information Systems from the Technical University of Munich.
\endbio

\bio{}
\textbf{Pablo Restrepo Henao} is an IT consultant for software engineering and machine learning at Netlight Consulting. He has worked as software engineer and technical lead in multiple companies and is currently pursuing his Master’s degree in Computer Science at the Technical University of Munich. His main research interest is the application of natural language processing techniques in the software engineering area. 
\endbio

\bio{}
\textbf{Parisa Yousefi} is a line manager and owner of architecture with Business Solution System (BSS) in Ericsson. Having a background as a developer, her interest are with AI/ML, platform related advances and new technologies as well as core software engineering principles and methodologies such as Agile, Test Driven development \& requirement engineering.
\endbio

\bio{}
\textbf{Tedi Juricic} is a technical quality assurance officer in functional testing within the Business Support Solution Charging and Billing unit at Ericsson. He is responsible for asserting the official quality stamp on the work packages as well as assuring the overall quality of the BSS product.

\endbio

\bio{}
\textbf{Jeannette Radduenz} is a quality and test manager within the Platform Management \& Testing Services unit at Allianz Technology. She is mainly responsible for planning, coordination, and control of test activities.
\endbio

\bio{}
\textbf{Carsten Wiecher} is a development engineer at KOSTAL Automobil Elektrik GmbH \& Co. KG. Since 2013, Carsten has been working in different development projects in the field of e-mobility. His main focus is in the area of software integration for complex electronic control units. Since 2018, Carsten is also a research associate at the IDiAL institute which is part of Dortmund University of Applied Sciences and Arts. Since 2020, he is involved in a research project with KOSTAL focusing on model-based systems engineering, requirements analysis and test specification. Carsten holds a Master´s degree in Information Technology from Dortmund University of Applied Sciences and Arts.
\endbio

\end{document}